\documentclass[useAMS,usenatbib]{mn2e}

\usepackage{defs}
\usepackage{epsfig}
\usepackage{graphicx}
\usepackage{amsmath}
\usepackage{amssymb}
\usepackage{lscape}
\usepackage{url}
\usepackage{setspace}
\usepackage{float}

\title[HerMES: Disentangling AGN and SF in radio sources]{HerMES: Disentangling active galactic nuclei and star formation in the radio source population} 
\author[J. I.~Rawlings et al.]
{\parbox{\textwidth}{\raggedright J. I.~Rawlings,$^{1}$\thanks{E-mail: \texttt{jason.rawlings.09@ucl.ac.uk}}
M. J.~Page,$^{1}$ M.~Symeonidis,$^{1,2}$ J.~Bock,$^{3,4}$ A.~Cooray,$^{5}$ D.~Farrah,$^{6}$ K.~Guo,$^{2}$ E.~Hatziminaoglou,$^{7}$ E.~Ibar,$^{8}$ S.J.~Oliver,$^{2}$ I.G.~Roseboom,$^{9}$ Douglas Scott,$^{10}$ N.~Seymour,$^{11,12}$ M.~Vaccari,$^{13,14}$ J.L.~Wardlow,$^{15}$}\vspace{0.4cm}\\
$^{1}$ Mullard Space Science Laboratory, University College London, Holmbury St. Mary, Dorking, Surrey, RH5 6NT, UK\\
$^{2}$ Astronomy Centre, Dept. of Physics \& Astronomy, University of Sussex, Brighton, BN1 9QH, UK\\
$^{3}$ California Institute of Technology, 1200 E. California Blvd., Pasadena, CA 91125, USA\\
$^{4}$ Jet Propulsion Laboratory, 4800 Oak Grove Drive, Pasadena, CA 91109, USA\\
$^{5}$ Department of Physics \& Astronomy, University of California, Irvine, CA 92697, USA\\
$^{6}$ Department of Physics, Virginia Tech, VA 24061, USA\\
$^{7}$ ESO, Karl-Schwarzschild-Str. 2, 85748 Garching bei M\"{u}nchen, Germany\\
$^{8}$ Instituto de F\'isica y Astronom\'ia, Universidad de Valpara\'iso, Avda. Gran Breta\~na 1111, Valpara\'iso, Chile\\
$^{9}$ Institute for Astronomy, Blackford Hill, Edinburgh EH9 3HJ, UK\\
$^{10}$ Department of Physics \& Astronomy, University of British Columbia, 6224 Agricultural Road, Vancouver, BC V6T 1Z1, Canada\\
$^{11}$ International Centre for Radio Astronomy Research, Curtin University, Perth, Australia\\
$^{12}$ CSIRO Astronomy \& Space Science, PO Box 76, Epping, NSW, 1710, Australia\\
$^{13}$ Astrophysics Group, Physics Department, University of the Western Cape, Private Bag X17, 7535 Bellville, Cape Town, South Africa\\
$^{14}$ INAF - Istituto di Radioastronomia, via Gobetti 101, 40129 Bologna, Italy\\
$^{15}$ Dark Cosmology Centre, Niels Bohr Institute, University of Copenhagen, Denmark}

\date{Accepted 2015 July 11.  Received 2015 June 28; in original form 2014 April 18}
\pagerange{\pageref{firstpage}--\pageref{lastpage}}
\pubyear{2015}

\begin{document}

\maketitle

\label{firstpage}

\begin{abstract}
We separate the extragalactic radio source population above $\sim$50\,$\mu$Jy into active galactic nuclei (AGN) and star-forming sources. The primary method of our approach is to fit the infrared spectral energy distributions (SEDs), constructed using {\it Spitzer}/IRAC and MIPS and {\it Herschel}/SPIRE photometry, of 380 radio sources in the Extended Chandra Deep Field-South. From the fitted SEDs, we determine the relative AGN and star-forming contributions to their infrared emission. With the inclusion of other AGN diagnostics such as X-ray luminosity, {\it Spitzer}/IRAC colours, radio spectral index and the ratio of star-forming total infrared flux to {\it k}-corrected 1.4\,GHz flux density, $q_{\rm IR}$, we determine whether the radio emission in these sources is powered by star formation or by an AGN. The majority of these radio sources (60 per cent) show the signature of an AGN at some wavelength. Of the sources with AGN signatures, 58 per cent are hybrid systems for which the radio emission is being powered by star formation. This implies that radio sources which have likely been selected on their star formation have a high AGN fraction. Below a 1.4\,GHz flux density of 1\,mJy, along with finding a strong contribution to the source counts from pure star-forming sources, we find that hybrid sources constitute 20$-$65 per cent of the sources. This result suggests that hybrid sources have a significant contribution, along with sources that do not host a detectable AGN, to the observed flattening of the source counts at $\sim$1\,mJy for the extragalactic radio source population.
\end{abstract}

\begin{keywords}
galaxies: active$-$galaxies: nuclei$-$galaxies: ISM$-$infrared: galaxies$-$radio continuum: galaxies
\end{keywords}

\section{Introduction}
In the time since the radio sky was first imaged, it has been found that extragalactic radio sources can be characterised by two distinct types. The first is where the production of the radio emission is a result of the accretion of material onto a super massive black hole at the centre of a galaxy \citep{Rees:84}, an active galactic nucleus (AGN). The second is where the emission stems from the production of stars on a massive scale \citep{Condon:92}, a star-forming galaxy (SFG). The bright end of the Euclidean-normalised source counts (151\,MHz flux densities\,$\sim$100\,mJy) is dominated by powerful radio-loud AGN \citep[e.g.][]{Willott:02}. The distinction between radio-loud and radio-quiet AGN was originally based on their 5\,GHz to $B$ band flux density ratios \citep{Kellermann:89}. Extending down to fainter flux densities, the source counts distribution is well described by a power law to a flux density of $\sim$1\,mJy, where the distribution then flattens \citep[e.g.][]{Windhorst:93}. The exact cause of this feature is unclear. There is evidence to suggest that the flattening is the result of the emergence of the SFG population, which therefore dominates the source counts below $\sim$1\,mJy \citep{Benn:93, Hopkins:98, Seymour:04, Condon:12}. However, there is increasing evidence that radio-quiet AGN could also have an important contribution at these faint flux densities \citep{Gruppioni:99, Ciliegi:03, Simpson:06, Seymour:08, Smolcic:08, Padovani:09} and therefore be partly responsible for the flattening.

To investigate the nature of the extragalactic radio population, observations at multiple wavelengths are required. The Extended {\it Chandra} Deep Field-South (ECDF-S) is a region that has been extensively observed at many wavelengths, e.g. in the radio \citep{Norris:06, Afonso:06, Kellermann:08, Miller:08}, infrared \citep{Lonsdale:03, Olsen:06, Taylor:09, Damen:11}, optical \citep{Rix:04, Le_Fevre:05, Hildebrandt:06, Cardamone:10} and X-ray \citep{Giacconi:02, Lehmer:05, Luo:08, Xue:11} and is therefore an ideal field for such a study. A vital component of a study such as this is the correct separation of emission from AGN and star-forming processes. This can be difficult because star formation diagnostics typically either reside at wavelengths where they suffer from obscuration (e.g. ultraviolet and optical) and/or contamination from AGN emission (e.g. X-ray, mid-infrared and radio). However, at far-infrared wavelengths, AGN can be weak \citep{Hatziminaoglou:10} and the thermal radiation from the cool dust associated with star formation can dominate. With the advent of the Spectral and Photometric Imaging REceiver \citep[SPIRE,][]{Griffin:10} instrument onboard the {\it Herschel Space Observatory}\footnote{{\it Herschel} is an ESA space observatory with science instruments provided by European-led Principal Investigator consortia and with important participation from NASA.} \citep{Pilbratt:10}, we now have a clear view of the Universe at these wavelengths, enabling us to probe the obscured star-forming properties of galaxies at various cosmic epochs. With the addition of {\it Herschel}/SPIRE data to {\it Spitzer} near- and mid-infrared data, infrared (IR) spectral energy distributions (SEDs) have been constructed for both IR-luminous AGN \citep[e.g][]{Hatziminaoglou:10, Seymour:11} and SFGs \citep[e.g.][]{Brisbin:10, Rowan-Robinson:10, Buat:10, Symeonidis:13}. Through multi-component SED fitting of AGN/SFG composite objects, previous studies have disentangled the contributions from these two processes to the IR emission of extragalactic populations \citep[e.g.][]{Mullaney:11, Feltre:13, Huang:14}. IR SED fitting can therefore be a powerful tool for detecting AGN, even in cases where they do not dominate the emission.

It is now well known that there is a tight correlation between the far-IR and radio emission produced from star formation \citep[the far-IR-Radio Correlation; FIRRC,][]{De_Jong:85, Helou:85}. A correlation with such small scatter, spanning many orders of magnitude in luminosity, may be surprising considering that the two forms of emission are generated by different processes. On one hand the IR emission is produced by the dust in a galaxy that absorbs the ultraviolet light from newly-formed massive stars and re-emits this energy as thermal radiation, while on the other, the radio emission is synchrotron radiation produced by the cosmic rays from supernova remnants. The reason for the correlation is that, as the life-cycle of massive stars is relatively short ($\sim$10$^{7}$\,yrs), the two processes can be observed simultaneously if star formation occurs continuously on longer timescales \citep{Ivison:10}. Many studies have calculated the logarithmic ratio of IR and radio luminosities, $q_{\rm IR}$, at both low redshift \citep[$z$\,$<$\,0.5,][]{Condon:91, Jarvis:10} and high redshift \citep[up to $z$\,$\sim$3,][]{Sajina:08, Ibar:08, Ivison:10, Chapman:10, Seymour:11}, either for its use in future studies or as a way of validating their results. The FIRRC can be a useful tool because through radio observations, the IR star-forming properties of a sample can be inferred (or vice versa). 

In this paper, we divide the radio source population into sources whose radio emission is powered by an AGN or by star formation, partly through fitting the IR SEDs of 380 radio sources in the ECDF-S. {\it Herschel}/SPIRE photometry, provided by the {\it Herschel} Multi-tiered Extragalactic Survey \citep[HerMES,][]{Oliver:12}, is included in the SEDs and we take advantage of the excellent multi-wavelength coverage of the field. From the fitted-SEDs, we constrain the AGN and galaxy interstellar medium (ISM) contributions to the IR emission and attempt to determine if the radio emission is powered by star formation or by an AGN. We then calculate the differential radio source counts for the ECDF-S by source type. The paper is structured as follows. In Section \ref{Section:sample}, we describe the master catalogue, multi-wavelength data and the sample selection. In Section \ref{Section:Contributions to radio}, we descibe the SED modelling and the fitting procedure. In Section \ref{Section:Results}, we present the results and discuss them in Section \ref{Section:Discussion}. Conclusions are presented in Section \ref{Section:Conclusion}. Throughout, we assume a $\Lambda$ Cold Dark Matter Universe, with $H_{\rm 0}$\,=\,70\,km\,s$^{-1}$\,Mpc$^{-1}$, $\Omega_{\rm \Lambda}$\,=\,0.7 and $\Omega_{\rm m}$\,=\,0.3.

\section{The sample}
\label{Section:sample}
\subsection{Master catalogue and photometric data}
\label{Section:Master catalogue and photometric data}
The master source catalogue comes from the second data release (DR2) from the National Radio Astronomy Observatory (NRAO) Very Large Array (VLA) observations of the ECDF-S at 1.4\,GHz \citep{Miller:13}. The image has a typical sensitivity of 7.4\,$\mu$Jy per 2.8\,by\,1.6\,arcsec$^2$ beam and an area of 0.32\,deg$^2$. The catalogue contains 883 sources above 5\,$\sigma$, with positional accuracies of 0.2\,arcsec and 0.3\,arcsec in right acension and declination, respectively, for a 5\,$\sigma$ source. Regarding the 1.4\,GHz flux densities, $S_{\rm 1.4}$, of the catalogue, it has been stated that the true radio flux density of an unresolved source is better represented by its peak flux density rather than by its integrated flux density \citep{Owen:08}. In line with other radio surveys, \citet{Miller:13} recommended the use of the peak flux densities for unresolved sources and the integrated flux densities for resolved sources. See \citet{Miller:13} for a description of how the radio sources were identified as resolved or unresolved. Through the use of multi-wavelength data and counterpart identifications \citep{Bonzini:12}, \citet{Miller:13} flagged 17 multiple-component sources likely to be powerful radio galaxies. For these sources, we used the summed flux densities of the individual components from the main catalogue. We also used the radio positions from the main catalogue and these corresponded to the sources that are likely to be the core objects. For the multiple-component sources for which the core is not visible, the average of the other component positions were given in the catalogue.

Some of the data used to obtain the IR SEDs come from the seven-band imaging of the CDFS carried out with the IRAC and MIPS {\it Spitzer} instruments. To obtain IRAC flux densities we used the IRAC/MUSYC Public Legacy Survey in the ECDF-S \citep[SIMPLE,][]{Damen:11} version 3.0 catalogue. The survey has sensitivities of 1.1, 1.3, 6.3 and 7.6\,$\mu$Jy, at 3.6, 4.5, 5.8, and 8.0\,$\mu$m, respectively, at 5\,$\sigma$. We cross-correlated the master radio catalogue with the SIMPLE catalogue using a 1.5\,arcsec search radius and found IRAC counterparts for 770 sources (out of 883). We deemed the radio sources without an IRAC counterpart to be undetected in the IRAC bands if its position was within the coverage of the SIMPLE survey. See Section \ref{Section:IR SED fitting} for how we dealt with non-detections in the IRAC bands. For flux densities at 24\,$\mu$m, we used the catalogue produced by the {\it Spitzer} far-Infrared Deep Extragalactic Legacy (FIDEL) Survey of the ECDF-S using IRAC priors \citep{Magnelli:09}. \citet{Magnelli:09} reported the sensitivity of the FIDEL map as 70\,$\mu$Jy at 5\,$\sigma$ and they presented a catalogue of $\sim$9400 sources above 3\,$\sigma$. Taking advantage of the fact that the FIDEL sources have already been assigned IRAC counterparts, we matched the IRAC SIMPLE counterparts to the FIDEL catalogue using the identical FIDEL IRAC positions. This gave 602 matches. For sources without 24\,$\mu$m counterparts, we used 24\,$\mu$m flux densities that were derived from the FIDEL image via the HerMES XID process \citep{Roseboom:10}. For radio sources with IRAC counterparts, flux densities were extracted at the IRAC positions, otherwise the radio positions were used as priors. For the radio sources with a 24\,$\mu$m counterpart, the XID 24\,$\mu$m flux densities were consistent with the flux densities from the \citet{Magnelli:09} catalogue. Furthermore, the mean flux density derived from the XID process of 0.50\,mJy is comparable to the mean flux density from the \citet{Magnelli:09} catalogue of 0.51\,mJy.

The remaining photometry comes from the MIPS 70\,$\mu$m and SPIRE 250, 350 and 500\,$\mu$m bands. The 70\,$\mu$m and SPIRE\footnote{The data presented in this paper will be released through the {\em Herschel} Database in Marseille HeDaM ({hedam.oamp.fr/HerMES}).} catalogues were generated by extracting sources from the 70\,$\mu$m FIDEL and HerMES maps of the CDF-S at the 24\,$\mu$m positions via the XID process. For the radio sources without a 24\,$\mu$m counterpart from the \citet{Magnelli:09} catalogue, the radio positions were used as priors. The noise in the 250, 350 and 500\,$\mu$m SPIRE bands is dominated by confusion noise and the images have 1\,$\sigma$ sensitivities of 5.8, 6.3 and 6.8\,mJy, respectively \citep{Nguyen:10}. The estimated uncertainties on the SPIRE flux densities take confusion noise into account.

\subsection{Sample selection and redshifts}
\label{Section:Sample selection and redshifts}
For consistency, we initially required coverage across all nine bands; IRAC 3.6, 4.5, 5.8, and 8.0\,$\mu$m, MIPS 24 and 70\,$\mu$m and SPIRE 250, 350 and 500\,$\mu$m. We therefore rejected the radio sources from the master catalogue that fell outside the IRAC, 24\,$\mu$m and/or SPIRE coverage; this removed 174 sources. We made further cuts by rejecting radio sources that had an IRAC counterpart that was flagged in the SIMPLE catalogue as being a Galactic star or had a flux density that was flagged as being contaminated by a Galactic star; this applied to 34 sources. We also rejected two radio sources that were matched to the same IRAC source and we rejected any radio sources that had a FIDEL 24\,$\mu$m source between 1.5 and 2.5 arcsec. This last rejection was on the basis that neighbouring sources are likely to be contaminating the far-IR flux densities. This removed 27 sources. After all these rejections we were left with a sample of 646 radio sources (hereafter called `full-sample'), each with photometry, including upper limits, across all nine IR bands. To estimate the probability of the radio sources being matched randomly to the counterpart sources, we shifted the radio positions of the master catalogue in right ascension by $\pm$\,10\,arcsec, while keeping the declination constant and then vice-versa. We cross-correlated the master catalogue again with the SIMPLE catalogue using the same search radius as used initially, this time with the four new positions. We found that the mean number of false matches for the four shifted positions is 83, 10 per cent of the matches for the true radio positions. We also did this with the 24\,$\mu$m catalogue, again using the same search radii, and found 16 false matches (3 per cent).

To fit model SEDs to the observed photometry, redshifts were necessary. \citet{Bonzini:12} identified the optical, IR and X-ray counterparts of the radio sources in the VLA ECDF-S DR2 catalogue. See their Table 4 for a list of the optical and IR catalogues. Their X-ray data come from the {\it Chandra X-ray observatory} 4\,Ms survey of the central part of the CDFS \citep[0.1\,deg$^2$,][]{Xue:11} and a mosaic of four 250\,ks observations of the extended field \citep[0.28\,deg$^2$,][]{Lehmer:05}. \citet{Bonzini:12} also acquired optical spectra for 13 radio sources which they combined with data from the literature to gain redshift information for as many of these objects as possible. For radio sources with both spectroscopic and photometric redshifts, if the spectroscopic redshift had a low quality flag, the photometric redshift was preferred by \citet{Bonzini:12}. They found the accuracy of the photometric redshifts to be $\sim$6 per cent. All sources with a low quality spectroscopic redshift also had a photometric redshift and so no source had a final redshift that is a low-quality spectroscopic redshift. Of the 646 radio sources in the full-sample, 567 (88 per cent) have redshifts, of which 220 (39 per cent) are spectroscopic and 347 (61 per cent) are photometric. 

\begin{figure}
\epsfig{file=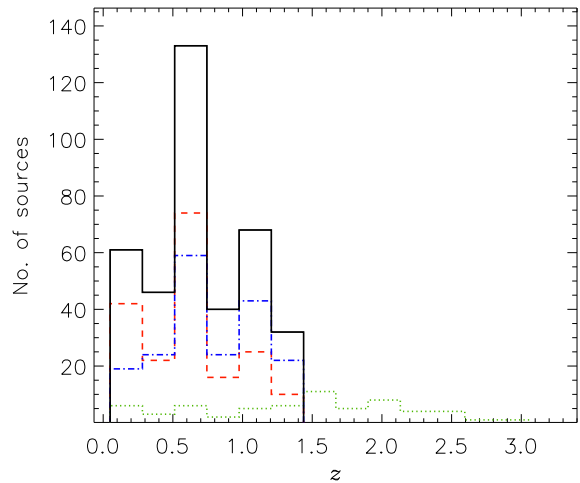, width=1\linewidth}
\caption{The redshift distribution of the sources in the sub-sample (for definition see text). The black solid, red dashed and blue dot-dashed lines show the total, spectroscopic and photometric redshift distributions, respectively. Also shown as a green dotted line is the redshift distribution of the sources excluded from the sub-sample because their photometric redshift uncertainties were too large to constrain the AGN and galaxy ISM SED components, see Section \ref{Section:IR SED fitting}.}
\label{Fig:z_dist}
\end{figure}

When it came to fitting the IR SEDs of the radio sources, a rest-wavelength range of 2$-$1000\,$\mu$m was used. For increasingly high$-$redshift sources, an increasing number of the IRAC bands are redshifted outside this wavelength range. This decreases the number of photometric points used during the fitting. To ensure there was a sufficient number of degrees of freedom to constrain the AGN and galaxy ISM SED components, we therefore also rejected sources above $z$\,$\sim$\,1.4. This removed 123 sources. Next the IR SEDs of the remaining 444 sources were fitted and the effect of redshift uncertainties for the sources with photometric redshifts was investigated (see Section \ref{Section:IR SED fitting} for a full description of the SED fitting procedure). We fitted the SED of each source three times; first using the redshift measurement and then using the redshift 3\,$\sigma$ Gaussian upper and lower bounds. Each best-fit returned a minimum $\chi^{2}$ value and the difference in these $\chi^{2}$ values from the best case to the worse case scenarios was measured. If the difference in $\chi^{2}$ was $>$ 9, the 3\,$\sigma$ confidence level for one `interesting parameter' when using a standard $\chi^{2}$ minimisation technique \citep{Press:92}, then the fits were not consistent at a significance of 3$\sigma$. This meant that the redshift uncertainty was too large to constrain the AGN and galaxy ISM SED components for any source that met this $\Delta\chi^{2}$ criterion. Here, the term ISM refers to the IR emission from a galaxy that is attributed to star formation i.e. cold dust and polycyclic aromatic hydrocarbon (PAH) emission. This criterion occured for 64 sources. We were therefore able to separate the AGN and galaxy ISM contributions to the IR emission for 380 of the 646 sources in the full-sample. These 380 radio sources constitute a sub-group of the full-sample (hereafter called `sub-sample'). The majority of the work presented here is based on this sub-sample, although we do consider the full-sample when calculating the radio source counts. The redshift distribution of the sub-sample peaks at $z$\,$\sim$0.6 and is displayed in Figure \ref{Fig:z_dist}. There are non-detections for four sources in the four IRAC bands and 26 sources at 24\,$\mu$m. There are X-ray counterparts for 130 sources. The full- and sub-sample selections are summarised in Table \ref{Table:sample}.

\subsection{Ancillary data}
\label{Section:Ancillary data}
Radio spectral indices were desired for the sub-sample and so we also determined 5.5\,GHz flux densities. To do this we used the catalogue of 145 sources from the ATLAS 5.5\,GHz survey of the EDCF-S \citep{Huynh:12}. The ATLAS image has an almost uniform sensitivity of $\sim$12\,$\mu$Jy rms, a beam size of 4.9\,arcsec\,$\times$\,2\,arcsec and an area of 0.25\,deg$^2$. We cross-matched the master catalogue to the ATLAS catalogue using the radio positions and a 1.5 arcsec search radius. There were ATLAS counterparts for 60 sources in the sub-sample. As with the master catalogue, for unresolved sources we used the peak flux densities and corresponding uncertainties for the 5.5\,GHz flux densities. For resolved sources, we used the integrated flux densities and assigned fractional uncertainties equal to the peak flux density fractional uncertainties.

\begin{table}
\begin{center}{
\caption{Breakdown of sources in the master radio catalogue (see Section \ref{Section:Sample selection and redshifts}). The full-sample refers to the sources from the master radio catalogue for which there was photometry, including upper limits, across all nine IR bands used in this study. The sub-sample refers to the sources in the full-sample for which we were able to separate the AGN and galaxy ISM contributions to the IR emission (see Section \ref{Section:IR SED fitting}).}
\begin{tabular}{@{}lc@{}}
\hline
\hline
Number of sources in master catalogue & 833\\
\hline
Number of sources in full$-$sample & 646\\
Number of sources in full$-$sample with redshifts & 567\\
Number of sources in sub$-$sample & 380\\
\hline
Number of sources in sub$-$sample with X$-$ray data & 130\\
\hline
\end{tabular}

\label{Table:sample}}
\end{center}
\end{table}

There were ten multiple-component sources in the ATLAS catalogue and the individual components were listed separately. As only a single central position was given for the multiple-component VLA sources, there were larger offsets than expected between the extended VLA sources and their ATLAS counterparts. Except for two cases, all multiple-component ATLAS sources were matched to multiple-component VLA sources. For one of these cases, the ATLAS source was interpreted to be two separate sources in the VLA catalogue and for the other case, because of the extended nature of the ATLAS source components, there was a separation larger than 1.5 arcsec between the ATLAS component positions and the VLA position. For the other eight multiple-component ATLAS sources, the individual ATLAS component flux densities were combined to give total 5.5\,GHz flux densities. The radio spectral index, $\alpha _{\rm rad}$, defined by $S_{\rm \nu}$\,$\propto$\,$\nu^{\alpha}$, was then calculated for the 60 sources with 5.5\,GHz counterparts using 1.4 and 5.5\,GHz flux densities. For the sources in the sub-sample without a 5.5\,GHz counterpart, we assumed a radio spectral index of $-$0.8.

For the sources with X-ray data from the \citet{Xue:11} and \citet{Lehmer:05} catalogues, we calculated rest-frame 0.5$-$8\,keV luminosities, $L_{\rm 0.5-8}$, using observed-frame 0.5$-$8\,keV fluxes, $f_{\rm 0.5-8}$. The equation we used was $L_{\rm 0.5-8}$\,=\,4$\pi\,D_{\rm L}^{2}\,f_{\rm 0.5-8}\,(1+z)^{\Gamma-2}$, where $\Gamma$ is the X-ray photon index and a value of 1.8 was assumed \citep{Xue:11}. Note these luminosities are not absorption-corrected, but because such a correction would only increase the luminosity, any sources identified as AGN through their X-ray luminosities (see Section \ref{Section:AGN indicators}) would remain identified as AGN if their luminosities were corrected. This could, however, mean that heavily-obscured AGN may not be identified through their X-ray emission.

\section{Determining the AGN and ISM contributions to the radio output} 
\label{Section:Contributions to radio}
\subsection{IR SED fitting}
\label{Section:IR SED fitting}
For the sub-sample of 380 sources, we determined the AGN and galaxy ISM contributions to their IR emission. To do this we used an IDL fitting procedure based on {\tt DecompIR}\footnote{https://sites.google.com/site/decompir} \citep{Mullaney:11}. We made certain asssumptions on the shapes of AGN and ISM SEDs though our choice of templates. We characterised the emission from the AGN using the \citet{Elvis:94} radio-quiet QSO SED and the {\it Spitzer} Wide-Area Infrared Extragalactic (SWIRE) template library \citep{Polletta:07}. We used five of the seven empirical AGN templates from the library and excluded the two Seyfert galaxy SEDs. These two Seyfert SEDs exhibit strong PAH features and therefore there is a non-negligible contribution from star formation in the host galaxies. To model the emission from the ISM we used the \citet[hereafter SK07]{Siebenmorgen:07} library of SEDs for starburst and ultraluminous IR galaxies. These were created using radiative transfer models and the library contains $\sim$7000 model SEDs that vary in total luminosity, starburst nucleus radius, visual extinction, ratio of the luminosity of OB stars embedded in dense clouds to the total luminosity and dust density in these clouds. \citet{Symeonidis:13} found that up to $z$\,$\sim$2, some of the SK07 SEDs were not representative of IR-luminous (log($L_{\rm IR}$/L$_{\rm \odot}$)\,$>$\,10) SFGs. They characterised the SK07 SEDs in terms of a `flux' parameter which was defined as $L_{\rm tot}$\,/\,4$\pi$\,$R^{2}$, where $L_{\rm tot}$ is the luminosity of the SED in units of $L_{\odot}$ and $R$ is the nuclear starburst radius in units of kpc. They found that the SEDs with log(flux\,/\,$L_{\odot}$\,kpc$^{-2}$)\,$\lesssim$\,8 or log(flux\,/\,$L_{\odot}$\,kpc$^{-2}$)\,$\gtrsim$\,11 were not good fits to their sample. Taking their results into account, we therefore considered only the SK07 SEDs with a flux in the range 8\,$<$\,log(flux\,/\,$L_{\odot}$\,kpc$^{-2}$)\,$<$\,11. Of the $\sim$7000 SEDs in the original library, 4765 met this criterion. By only using these SEDs as the templates for our ISM component, we have assumed that the IR ISM SEDs of our sub-sample are similar to those of IR-luminous SFGs. If the IR emission of our sources is star-formation-dominated then this would imply they are also IR-luminous SFGs and this assumption is therefore reasonable. If, on the other hand, these sources are quiescently forming stars then they could have ISM SEDs that are different to those of IR-luminous SFGs. They could perhaps have SEDs more like the cirrus-dominated SEDs observed by \citet{Rowan-Robinson:10}. However, if they are quiescently forming stars then the AGN will likely be the dominant component in their near- to mid-IR SED. This then means that the disputed nature of an ISM component that has only a small contribution would not be a major issue. For consistency, we therefore fitted the 4765 SK07 SEDs to all sources in our sub-sample.

At near-IR wavelengths, there is a contribution to the total output of a galaxy from direct starlight. Although direct starlight is taken into account in the radiative transfer models for the SK07 templates, we desired an independent starlight component to increase the flexibility of our SED-fitting model. We therefore included a third component to represent this emission. For this component, we made use of the \citet{Bruzual:03} stellar population (SP) library. The library contains 78 templates which vary in star-forming history, stellar age (from 5\,Myr to 12\,Gyr) and metallicity (from $Z$\,=\,0.008 to 0.05) and cover the wavelength range from 9\,nm to 160\,$\mu$m. While the SEDs vary greatly at the shorter wavelengths, longward of $\sim$2\,$\mu$m they are all well approximated by a Rayligh-Jeans tail. Therefore, by only fitting at wavelengths $\gtrsim$\,2\,$\mu$m (rest-frame), we were able to use just one SP template as the starlight component. With the addition of this third component to our model, the direct light from the SK07 templates needed to be removed. This was done by subtracting the SP template, having first normalised it to the SK07 template flux at 2\,$\mu$m. These modified SK07 templates then purely represent the emission from the ISM of the galaxy. The normalisation of the SP template was fixed only when removing the direct starlight from each SK07 template. The SP component normalisation was a free parameter during the SED fitting.

To account for the reddening of the AGN that can occur in some sources we made an assumption that the dusty environments of high-redshift AGN can be modelled by a Galactic extinction curve \citep{Sajina:07a, Lacy:07, Seymour:08, Drouart:12, Rawlings:13}. We chose the \citet{Draine:03} extinction curve that uses $R_{V}$\,=\,3.1, the typical value for the Milky Way Galaxy, where $R_{V}$ is the ratio of total extinction, $A_{\rm V}$, to selective extinction, $A_{B}-A_{V}$, in the $V$ band (i.e. $R_{V}$\,=\,$A_{\rm V}$/($A_{B}-A_{V}$)) and we applied this curve to the AGN component only. Using such a curve could result in an AGN component with a large silicate absorption feature. This approach assumes that the obscuring medium acts as a simple foreground screen and is close to the AGN but not located in the torus itself. Through their clumpy torus modelling, \citet{Nenkova:08b} showed that the silicate absorption strength would be weak ($\tau_{\rm 9.7\,\mu m}$\,$<$\,1)\footnote{Throughout the paper $\tau_{\rm 9.7\,\mu m}$ represents the 9.7\,$\mu$m apparent optical depth and is defined as the natural log of the ratio of continuum emission to observed emission.} if the dust causing the AGN reddening was situated in the torus. Therefore, to also consider this scenario, we reapplied the extinction curve with the silicate feature removed by interpolating over the relevant wavelength region. This allowed for the case of an AGN component with a large amount of reddening but no silicate feature. There was no need to apply extinction to the ISM component because extinction was already applied to the models used to create the SK07 templates. We did not apply extinction to the direct starlight component because this would decrease the slope of the template and this component would start to become degenerate with the AGN component over the wavelength region in which the starlight component most influences the SED-fitting ($\lambda$\,$<$\,10\,$\mu$m rest-frame). This would have caused the direct starlight component to become redundant.

During the IR SED fitting we used six free parameters: (i) the chosen ISM template, (ii) the ISM component normalisation, (iii) the chosen AGN template, (iv) the AGN component normalisation, (v) the direct starlight component normalisation and (vi) the extinction of the AGN component i.e. $A(\lambda)$. An upper limit restriction was placed on $A(\lambda)$ such that $A_{\rm V}$\,$\sim$\,30 mag. This limit corresponds to a large 9.7\,$\mu$m silicate absorption depth of $\tau_{\rm 9.7\,\mu m}$\,$\sim$2 for the extinction curve used. A silicate absorption depth of this size has been observed in galaxies selected to be dusty i.e. large mid-IR-optical colours \citep[e.g.][]{Sajina:07a}. Only six free parameters have been listed here and so we make the caveat that there are other parameters embedded within the `choice of ISM template' parameter. These are the free parameters of the radiative transfer modelling which produced the SK07 SED library. We fitted all three components of our model in our rest-wavelength range and the expression for the model, $S_{\rm \nu}$, in mJy was
\\
\begin{equation}
{S_{\rm \nu}(\lambda)=k_{\rm a}a_{\rm \nu}(\lambda)\,10^{-A(\lambda)/2.5}+k_{\rm b}b_{\rm \nu}(\lambda)+k_{\rm c}c_{\rm \nu}(\lambda)}
\end{equation}
\\
{\noindent}where $k_{\rm a}$ and $a_{\rm \nu}(\lambda)$ are the normalisation factor and flux density of the AGN component respectively, $k_{\rm b}$ and $b_{\rm \nu}(\lambda)$ are the normalisation factor and flux density of the ISM component and $k_{\rm c}$ and $c_{\rm \nu}(\lambda)$ are the normalisation factor and flux density of the stellar population component. For each ISM template, $\chi^{2}$ minimisation was performed on the SED, simultaneously solving for $k_{\rm a}$, $k_{\rm b}$, $k_{\rm c}$ and $A(\lambda)$. The photometric data were assigned a minimum uncertainty of 10 per cent so that individual data points were not weighted too high in the $\chi^{2}$ estimation during the fitting. The minimum error is justified by the need to take into account the limitations of the templates which are being used in the modelling. Both the AGN and the starburst model components are taken from libraries of discrete SEDs rather than continuous parameter-based models, and the only degrees of freedom allowed beyond the choice of the library template are the normalisations of the templates and the degree of dust reddening of the AGN. SK07 compare their library SEDs to a number of well know star-forming galaxies, and show the residuals of their fits in their figure 7. The scatter shown in the various panels of that plot typically exceeds 10 per cent, though the statistical and systematic uncertainties on the data points contribute as well as the limitations of the models. Bearing that in mind, 10 per cent appears to be a reasonable estimate for the data-model mismatch which is likely to come from the limited set of SK07 SED shapes. In practical terms, such an approach is required because otherwise the statistical weight of the near-IR bands would be much greater than the far-IR bands, and the fitting of the near-IR photometry would dominate the goodness of fit. Without the adjustment to the photometric uncertainties at short wavelengths, we found that the best-fitting templates often fail to reproduce the far-IR emission, and therefore the SED fitting process would not provide reliable constraints on the ISM component. We note that in adopting a minimum error of 10 per cent we are following the same approach as that taken by \citet{Symeonidis:08a} and \citet{Symeonidis:08b} to model the SEDs of infrared galaxies using the SK07 templates.

We ensured that source non-detections in the IRAC bands had less weight than detections during the SED fitting. Our technique was to assign flux densities and 1\,$\sigma$ uncertainties that were equal to half the 5\,$\sigma$ survey limits of the SIMPLE survey, giving a S/N\,=\,1. For example, sources not detected at 8\,$\mu$m were given an 8\,$\mu$m flux density and uncertainty of 7.6\,$\mu$Jy/2\,=\,3.8\,$\mu$Jy. With this technique, at each wavelength in the 3.6$-$8\,$\mu$m range, non-detections had individual $\chi^{2}$ values of $<$\,1 for fitted models that predicted a flux density below the survey sensitivity. This meant such a model was acceptable at the 1\,$\sigma$ level at each wavelength where there was a non-detection. If a model predicted a flux density above the survey sensitivity then a non-detection would have an individual $\chi^{2}$ value of $>$\,1 and the model would not be acceptable at the 1\,$\sigma$ level.

To determine the confidence in the best-fitting ISM component normalisation ($k_{\rm b}$), we found the extent to which it can vary until the condition $\chi^{2}$\,$-$\,$\chi^{2}_{\rm min}$\,=\,$\Delta\chi^{2}$\,=\,$a^{2}$ is met, letting $a$\,=\,1 or 3 to find the 1 and 3\,$\sigma$ $\chi^{2}$ confidence levels for one `interesting parameter'. During this process the free parameters: $k_{\rm a}$, $k_{\rm c}$ and $A(\lambda)$ were allowed to vary also. Total IR (8$-$1000\,$\mu$m) star-forming fluxes, $S_{\rm IR,SF}$, were derived from the best-fitting ISM components. These were converted into luminosities, $L_{\rm IR,SF}$ and then into SFRs via the \citet{Kennicutt:98} relation
\\
\begin{equation}
\frac{\rm SFR}{\rm 1\,M_{\odot}\,yr^{-1}}\,=\,\frac{L_{\rm IR,SF}}{\rm 5.8\,\times\,10^{9}\,L_{\odot}}
\end{equation}
\\
{\noindent}The relation assumes solar abundances and a \citet{Salpeter:55} initial mass function (IMF). Uncertainties in $L_{\rm IR,SF}$ and the SFRs were derived from the uncertainties in the ISM normalisations. As $A(\lambda)$ was allowed to vary throughout, the uncertainties in the above parameters have also taken into account the uncertainty in the level of extinction applied. For example, where the extinction is not well constrained, the confidence levels on the above parameters include AGN components both with and without extinction. 

For each source, we also sought to determine if AGN and ISM components were required in the fitted SEDs. To do this we applied a more conservative significance cut than used previously as this determination will have important consequences later on for this study. Firstly, the reduction in the minimum $\chi^{2}$ as a result of the inclusion of an AGN component with respect to a purely star-forming template was measured. If this $\Delta\chi^{2}$ was $>$\,25 then at the 5\,$\sigma$ confidence level for one `interesting parameter' when using a standard $\chi^{2}$ minimisation technique \citep{Press:92}, the fits with and without the component were not consistent at a significance of 5$\sigma$. For any source that met this $\Delta\chi^{2}$ criterion we therefore determined that at the 5\,$\sigma$ level, an AGN component was deemed to be required in the IR SED, i.e. an AGN had been detected. Similarly, if the inclusion of an ISM component reduced the minimum $\chi^{2}$ by $>$\,25 with respect to a model consisting of only an AGN component, then an ISM component was deemed to be required at 5\,$\sigma$.

\subsection{AGN indicators}
\label{Section:AGN indicators}
For the 380 radio sources in the sub-sample we aimed to determine two characteristics: (1) whether or not an AGN is present in the galaxy; and (2) whether or not the observed radio emission is powered by an AGN. To determine characteristic (1), we made use of multi-wavelength data, as different regions of the AGN structure can be probed by their emission at different wavebands. Along with fitting the IR SEDs with AGN and ISM components, we also applied four observational-based indicators to each source and identified that an AGN is present if any of the following criteria were met: 

\begin{itemize}
\item[] (i) an X-ray luminosity greater than 10$^{42}$\,erg\,s$^{-1}$.
\item[] (ii) IRAC colours that satisfy the \citet{Donley:12} criteria for having an AGN-dominated near- to mid-IR SED.
\item[] (iii) a radio spectral index with a 3\,$\sigma$ upper bound\,$<$\,$-$1 or a 3\,$\sigma$ lower bound\,$>$\,$-$0.5.
\item[] (iv) if the sources resides more than 3\,$\sigma$ above the FIRRC. 
\end{itemize} 

For criterion (i), an X-ray luminosity cut of 10$^{42}$\,erg\,s$^{-1}$ is often used by extragalactic studies to differentiate between AGN and SFGs \citep[e.g. see][]{Fabbiano:89, Moran:96, Brandt:05}. However, this is not a theoretical upper limit to star-forming X-ray luminosities and so we make the caveat that it is possible for a small number of extreme star-forming sources in the sub-sample to have an X-ray luminosity above this cut. 

For criterion (iii), this spectral index range was used since it has been shown that synchrotron radiation at radio wavelengths, as a result of star-forming processes, has a spectrum with an index in the range $-$1.0 to $-$0.5 \citep{Thompson:06}. Although AGN could have radio spectral indices within this range, if a source has an index outside of this range then we infer that this is because of the presence of an AGN.

For criterion (iv), we calculated $q_{\rm IR}$ for each source using a slightly modified form of the equation first defined by \citet{Helou:85}. Following other studies \citep{Ivison:10, Chapman:10}, the equation we used was
\\
\begin{equation}
{q_{\rm IR}\,=\,{\rm log_{10}}\,\left[\frac{S_{\rm IR,SF}/3.75\,\times\,10^{12}\,{\rm W\,m^{-2}}}{S_{\rm 1.4\,GHz}/{\rm W\,m^{-2}\,Hz^{-1}}}\right]}
\end{equation}
\\
{\noindent}Radio 1.4\,GHz flux densities were {\it k}-corrected using the radio spectral indices we have derived (see Section \ref{Section:Ancillary data}) for sources with 5.5\,GHz counterparts, or otherwise using the assumed index of $-$0.8 \citep{Ivison:10}. If the $q_{\rm IR}$ value was less than the 3\,$\sigma$ lower bound of the median $q_{\rm IR}$ derived by \citet{Ivison:10}, $q_{\rm IR}$\,=\,2.40\,$\pm$\,0.24, then we inferred that the radio emission is produced by an AGN. If none of the AGN criteria (i$-$iv) was met and an AGN component was not required in the fitted SED then we have been unable to show that an AGN is present in the source. 

Next we determined characteristic (2), i.e. whether or not the observed radio emission is powered by an AGN, for the 380 sources in the sub-sample. If we have been unable to show that an AGN is present in the source then we inferred the radio emission of the source is powered by star formation. If there is an indication of jet activity in the source (i.e. criteria iii or iv) then we inferred the radio emission of the source is powered by an AGN, regardless of the results from the other indicators. If there is an indication of black hole accretion (i.e. criteria i or ii or through the SED fitting) but there is no sign of jet activity then the source is referred to as a hybrid. We use this nomenclature for the latter source type because the use of multi-wavelength data has revealed that an AGN is present in the source. However, as there is no sign of jet activity, it is not clear whether or not the radio emission is powered by the AGN.

\section{Results}
\label{Section:Results}
\begin{figure}
\epsfig{file=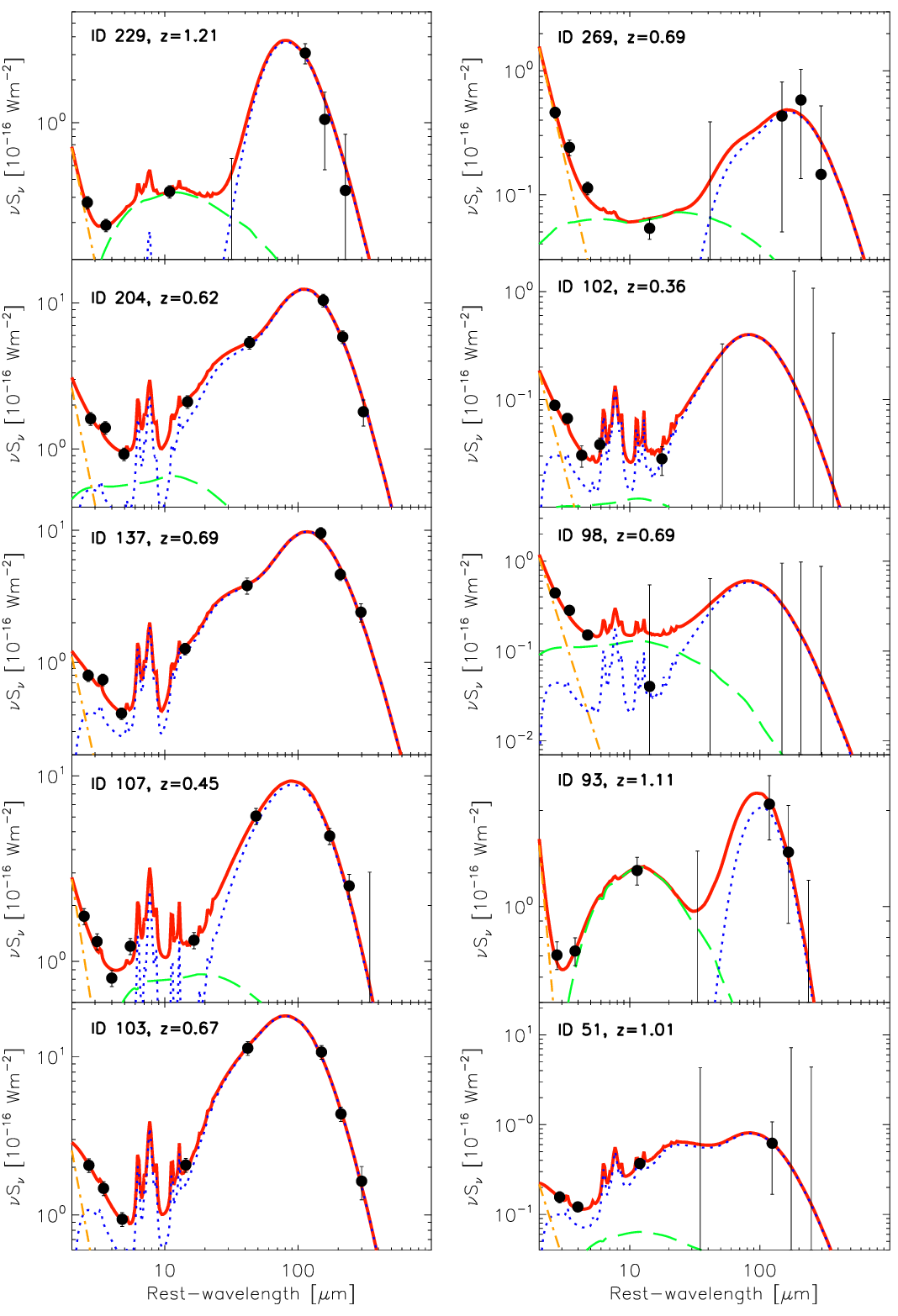, width=0.92\linewidth}
\caption{Examples of best-fitting models, denoted by a solid red line, for IR SEDs that, at $\ge$\,5\,$\sigma$, require an ISM component (left panels) and do not require an ISM component (right panels). Included are the separate components: AGN-green dashed line; ISM-blue dotted line; and direct starlight-orange dot-dashed line. Upper limits are 3\,$\sigma$.}
\label{Fig:sed_fitted_host}
\end{figure}

\begin{figure}
\epsfig{file=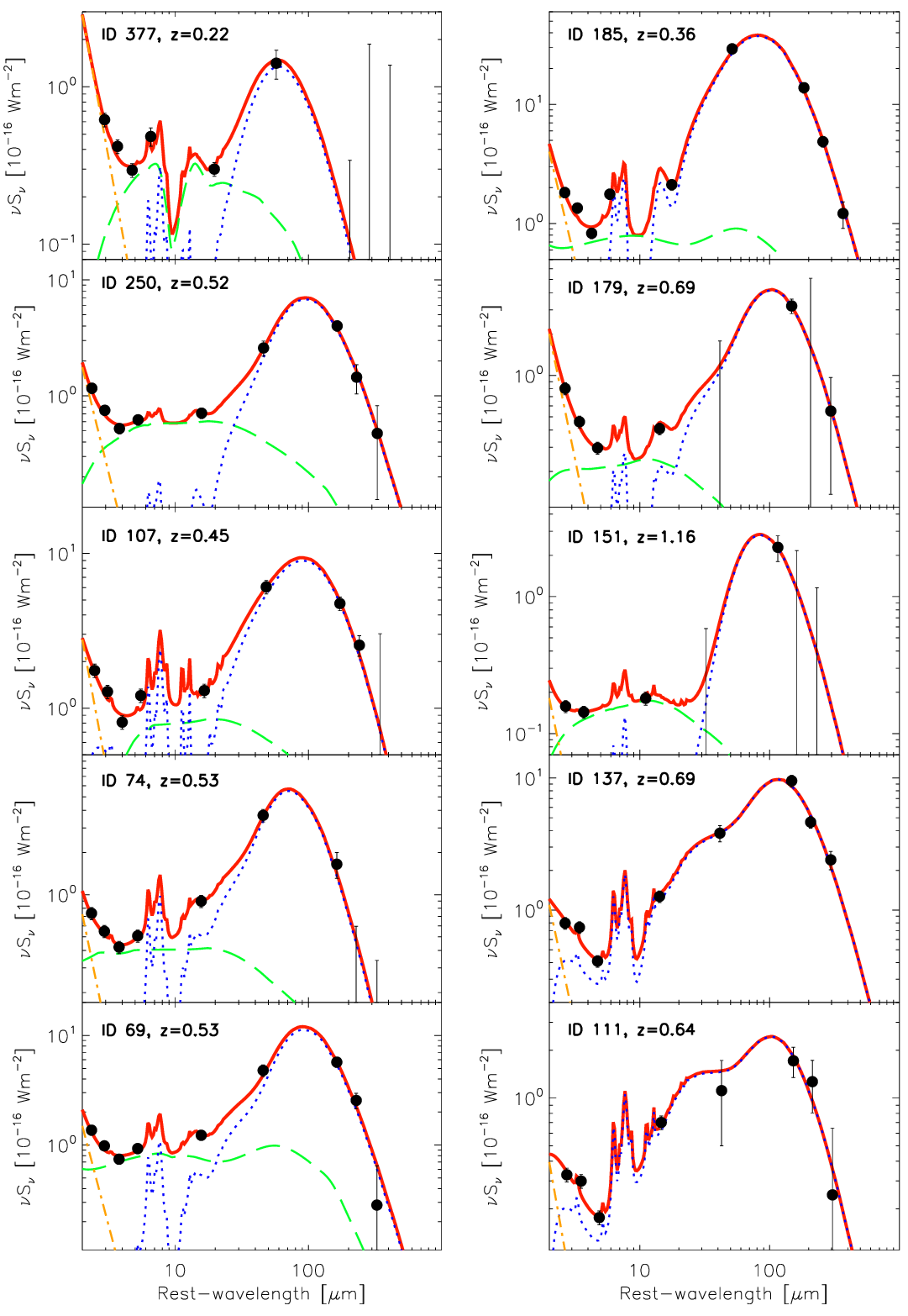, width=0.92\linewidth}
\caption{Examples of best-fitting models, denoted by a solid red line, for IR SEDs that, at $\ge$\,5\,$\sigma$, require an AGN component (left panels) and do not require an AGN component (right panels). Included are the separate components: AGN-green dashed line; ISM-blue dotted line; and direct starlight-orange dot-dashed line. Upper limits are 3\,$\sigma$.}
\label{Fig:sed_fitted_agn}
\end{figure}

Of 380 sources in the ECDF-S with fitted IR SEDs, the majority (227; 60 per cent) are identified as an AGN through their SED and/or by at least one of the observational-based indicators. For 253 of the 380 sources in the sub-sample, an ISM component is required in their fitted SEDs at the 5\,$\sigma$ level ($\Delta\chi^{2}$\,$>$\,25). All but five of these sources are detected in at least one SPIRE band at 3\,$\sigma$. For the remaining 127 sources, an ISM component is not required at the 5\,$\sigma$ level and thus we do not significantly detect ongoing star formation in these sources. For 123 of the 380 sources, an AGN component is required at the 5\,$\sigma$ level. All of these sources are detected in at least two IRAC bands at 3\,$\sigma$ but only 15 have an AGN fraction of the total IR luminosity that is $>$\,50 per cent. The fitted SEDs of the 123 sources have a median reduced $\chi^{2}$ of 6.4. The ISM templates cannot reproduce the high near- to far-IR ratios observed for these sources and so an additional AGN component significantly improves the fit. For 92 sources, neither an ISM component nor an AGN component is required. This is a result of the quality of the photometric data being too low to be able to contrain the contribution from each component. In Figures \ref{Fig:sed_fitted_host} and \ref{Fig:sed_fitted_agn}, we show several examples of best-fitting models to the IR SEDs for the sub-sample. In the right-hand panels of Figures \ref{Fig:sed_fitted_host} and \ref{Fig:sed_fitted_agn}, we show examples of SEDs that do not require an ISM and AGN component, respectively, at 5\,$\sigma$. However, this does not mean these components are not present in the best-fitting models.

The number of sources identified as an AGN decreases with increasing number of indicators considered and only two sources are identified as an AGN by all four indicators and through their SED. The indicators that find the highest and lowest AGN fractions are the X-ray luminosity (67/130; 52 per cent) and IRAC colours (15/380; 4 per cent) indicators, respectively. Of the 123 sources identified as AGN through their SED, 84 sources were not identified as an AGN by any of the indicators. All SEDs shown in the left panels of Figure \ref{Fig:sed_fitted_agn} are examples of the fitted SEDs of sources for which an AGN component was required at 5\,$\sigma$ but none of the indicators identified an AGN. The results for the various indicators are displayed in Table \ref{Table:AGN_identified}. The results are displayed in matrix form which allows the results for different combinations of indicators to be discerned. Overall, we find that for 153 of the 380 sources (40 per cent), the radio emission is powered by star formation (hereafter called `SF-powered' sources). For 95 sources (25 per cent), the radio emission is powered by an AGN (hereafter called `AGN-powered' sources). The remaining 132 sources (35 per cent) have been found to be hybrids\footnote{See Section \ref{Section:AGN indicators} for the SF-powered, AGN-powered and hybrid criteria.}.

\begin{table}
\begin{center}{
\caption{Number of sources identified as AGN by the various indicators. Column labels denote the indicator labels described in Section \ref{Section:AGN indicators}.}
\begin{tabular}{@{}ccccccc@{}}
\hline
\hline
Indicator & X-ray & IRAC & $\alpha$ & FIRRC & SED-fitting\\
& (i) & (ii) & (iii) & (iv) &\\
\hline
X-ray          & 67 & $-$ & $-$ & $-$ & $-$\\ 
IRAC           & 11 & 15 & $-$ & $-$ & $-$\\
$\alpha$    & 5   & 2 & 13 & $-$ & $-$\\
FIRRC          & 23 & 6 & 12 & 94 & $-$\\
SED-fitting & 28 & 13 & 5 & 18 & 123\\
\hline
Total           & 67 & 15 & 13 & 94 & 123\\
\hline
\end{tabular}

\label{Table:AGN_identified}}
\end{center}
\end{table}

\begin{figure*}
\epsfig{file=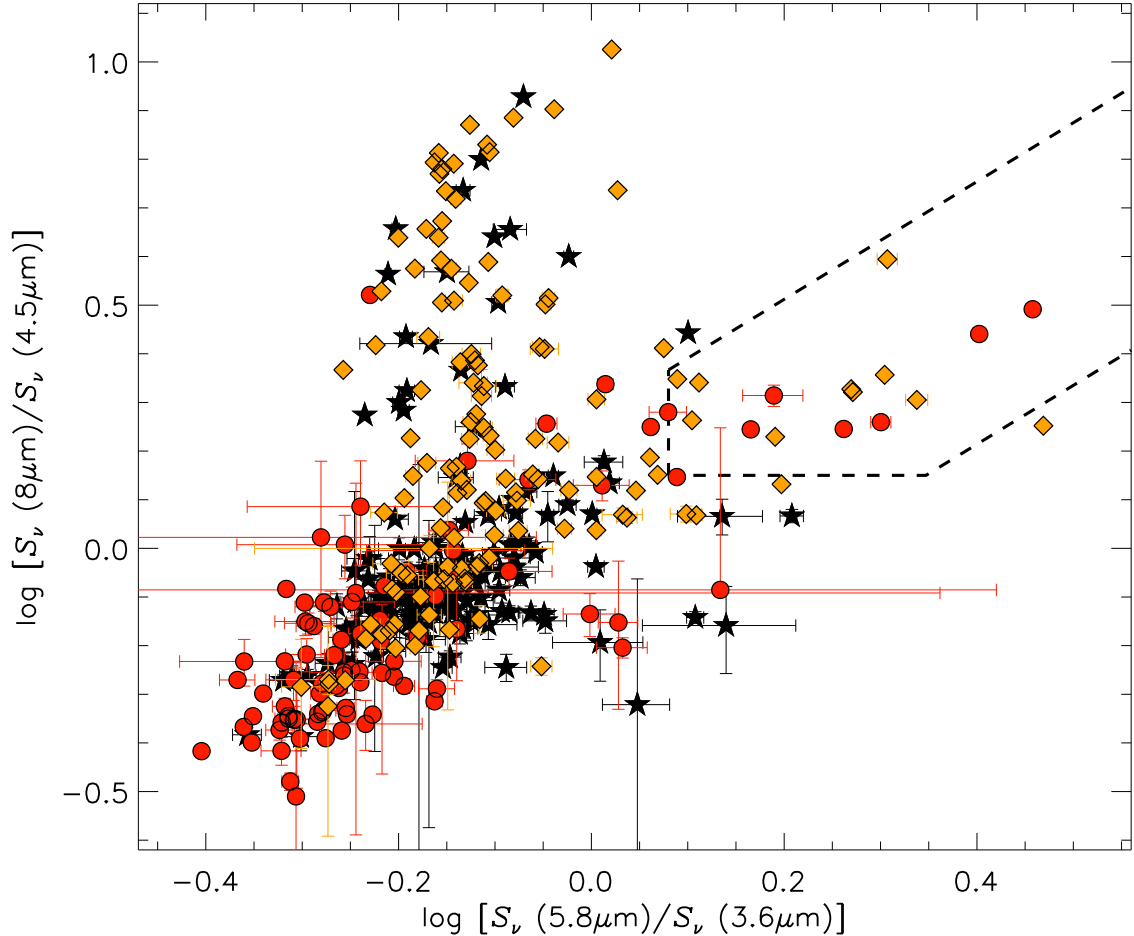, width=0.93\linewidth}
\caption{IRAC colours for the sub-sample. The area enclosed by the dashed lines is the region in colour space in which sources that have an AGN-powered near- to mid-IR SED reside, according to the \citet{Donley:12} criteria. SF-powered, AGN-powered and hybrid sources are shown as black stars, red circles and orange diamonds, respectively. Sources with upper limits to their IRAC flux densities (see Section \ref{Section:Master catalogue and photometric data}) lie off the plot.}
\label{Fig:irac_colours}
\end{figure*}

We next consider the fluxes, in the sense employed by \citet{Symeonidis:13} to characterise the SK07 templates, of the best-fitting ISM components for the different types of sources. We find that the ISM components of the SF-powered sources have comparable fluxes ($\langle$log(flux\,/\,$L_{\odot}$\,kpc$^{-2}$)$\rangle$\,=\,8.83, standard error\,=\,0.06) to those of the AGN-powered ($\langle$log(flux\,/\,$L_{\odot}$\,kpc$^{-2}$)$\rangle$\,=\,8.71, standard error\,=\,0.09) and hybrid ($\langle$log(flux\,/\,$L_{\odot}$\,kpc$^{-2}$)$\rangle$\,=\,8.81, standard error\,=\,0.07) sources. Although the 380 radio sources in our sub-sample have been distinguished in terms of whether an AGN is detected (at any wavelength), we find that there is no apparent difference in the shapes of their IR ISM SEDs. Overall, the majority of sources have star-forming total IR luminosities that would qualify them as either LIRGs or normal IR galaxies (NIRGs; log($L_{\rm IR}$/L$_{\rm \odot}$)\,$<$\,11). ULIRGs are less common and there are no HyLIRGs. Table \ref{Table:irgs} shows the breakdown of sources into the luminosity categories. Translating star-forming IR luminosity into SFR, there are some high-redshift sources with extreme SFRs ($\gtrsim$\,1000\,M$_{\rm \odot}$\,yr$^{-1}$). Comparable SFRs have been observed in high-redshift {\it Herschel}-selected galaxies \citep[e.g.][]{Rodighiero:11, Elbaz:11} and other non-{\it Herschel}-selected galaxies (e.g. SFGs, \citealt{Seymour:08}; high-redshift radio galaxies, \citealt{Rawlings:13}; and sub-millimeter galaxies, \citealt{Pope:06}). For all 380 sources, we find a median SFR of 20\,M$_{\rm \odot}$\,yr$^{-1}$, with a median absolute deviation (MAD) of 20\,M$_{\rm \odot}$\,yr$^{-1}$. For the SF-powered and AGN-powered sources, we derive median SFRs of 40\,(MAD\,=\,25)\,M$_{\rm \odot}$\,yr$^{-1}$ and 5\,(MAD\,=\,5)\,M$_{\rm \odot}$\,yr$^{-1}$, respectively, while for the hybrids, we find a median SFR of 20\,(MAD\,=\,15)\,M$_{\rm \odot}$\,yr$^{-1}$. At low redshift, the fraction of sources powered by star formation that host an AGN (i.e. hybrid/(hybrid+SF-powered)) is high: 0.7 at $z$\,$<$\,0.5. This fraction decreases at high redshifts to 0.4 for 0.5\,$<$\,$z$\,$<$\,1 and 0.3 for 1\,$<$\,$z$\,$<$\,1.4.

\begin{table}
\begin{center}{
\caption{Number of NIRGs, LIRGs and ULIRGs in the sub-sample according to their star-forming total IR luminosity.}
\begin{tabular}{@{}cccc@{}}
\hline
\hline
IR luminosity range & Type & $\#$ & Fraction\\
& & & [$\%$]\\
\hline
log($L_{\rm IR,SF}$/L$_{\rm \odot}$)\,$<$\,11 & NIRG & 162 & 43\\
11\,$\leq$\,log($L_{\rm IR,SF}$/L$_{\rm \odot}$)\,$<$\,12 & LIRG & 203 & 53\\
12\,$\leq$\,log($L_{\rm IR,SF}$/L$_{\rm \odot}$)\,$<$\,13 & ULIRG & 15 & 4\\
\hline
\end{tabular}
\label{Table:irgs}}
\end{center}
\end{table}

\begin{figure*}
\epsfig{file=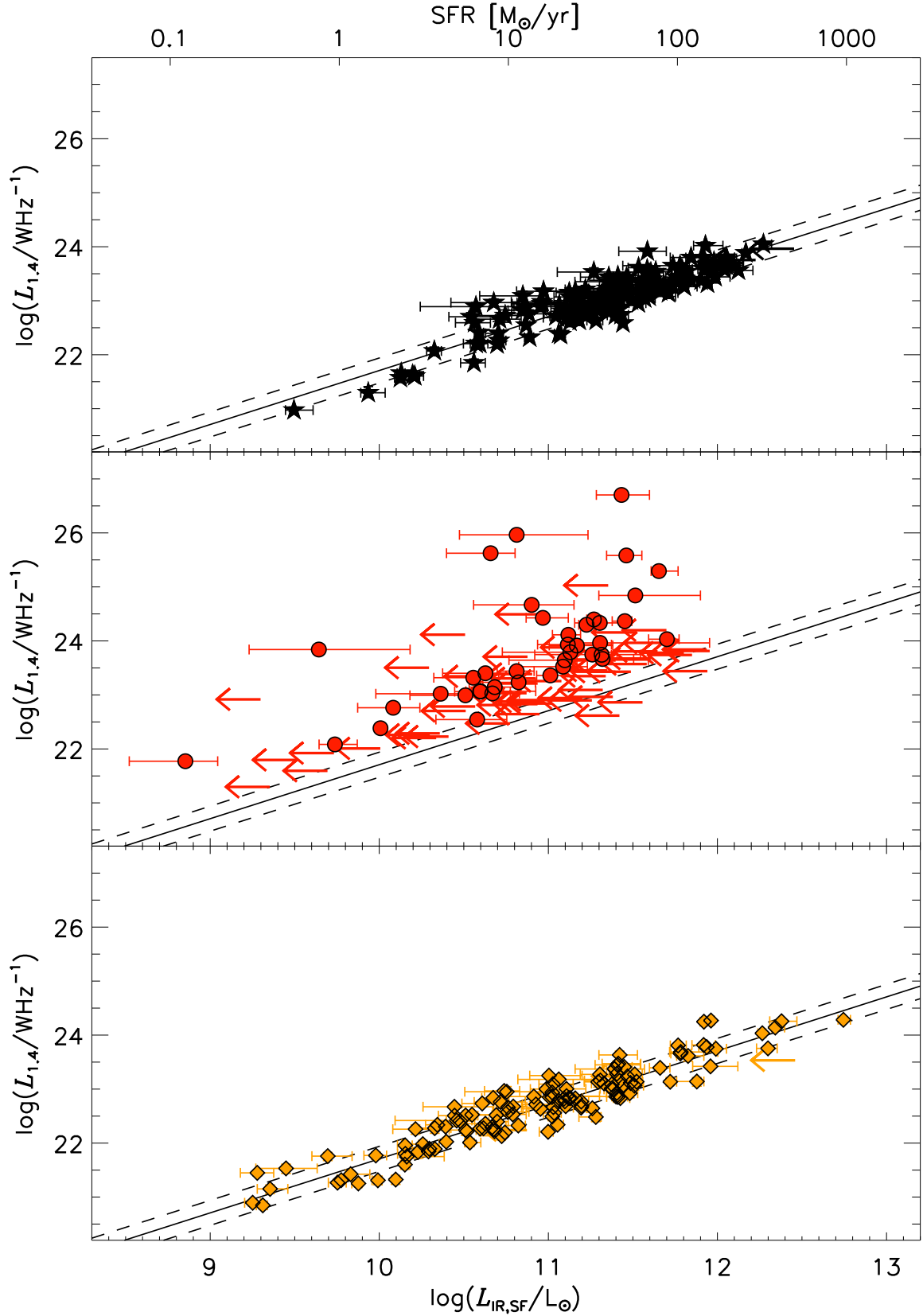, width=0.93\linewidth}
\caption{Rest-frame 1.4\,GHz luminosity as a function of star-forming IR luminosity (lower abscissa) and SFR (upper abscissa) for SF-powered (top panel, black stars), AGN-powered (middle panel, red circles) and hybrid (bottom panel, orange diamonds) sources. For sources with a $<$\,1.5\,$\sigma$ detected star-forming IR luminosity, 3\,$\sigma$ upper limits are shown. The solid line represents the FIRRC with a $q_{\rm IR}$ value equal to the mean SF-powered $q_{\rm IR}$ of 2.30, while the dashed lines represent the $\pm$\,1\,$\sigma$ limits, $\sigma_{q, \rm SF}$\,=\,0.23, of the correlation.}
\label{Fig:lrad_v_lir}
\end{figure*}

\begin{figure*}
\epsfig{file=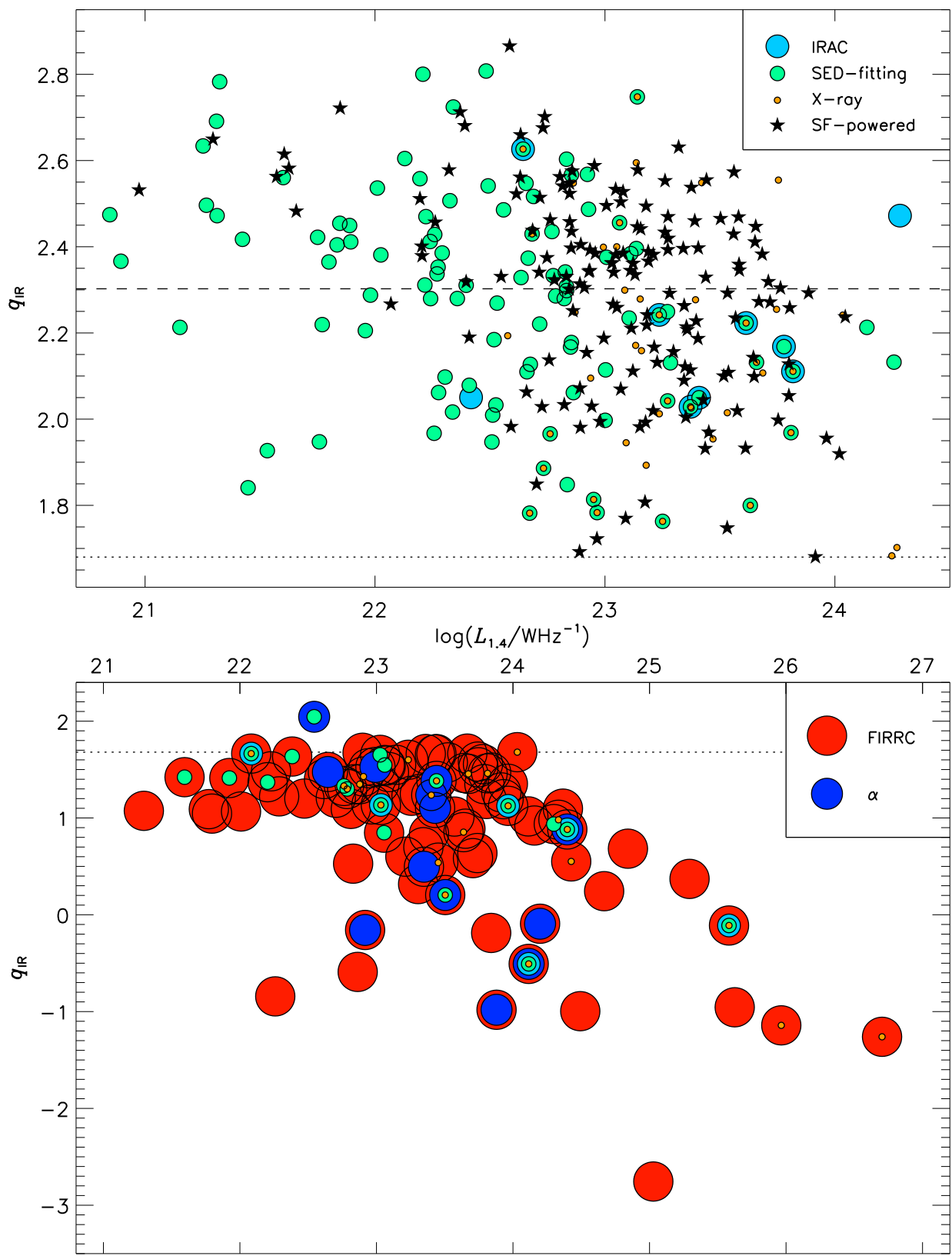, width=0.99\linewidth}
\caption{Value of $q_{\rm IR}$ as a function of 1.4\,GHz luminosity for SF-powered and hybrid sources (top panel) and AGN-powered sources (bottom panel). For each source, we show which indicators flagged an AGN with different coloured circles. The dashed line in the top panel represents the mean $q_{\rm IR}$ of the SF-powered sources and the dotted line in both panels identifies the region below which sources are more than 3\,$\sigma$ above the FIRRC.}
\label{Fig:AGN_identified}
\end{figure*}

\begin{figure*}
\epsfig{file=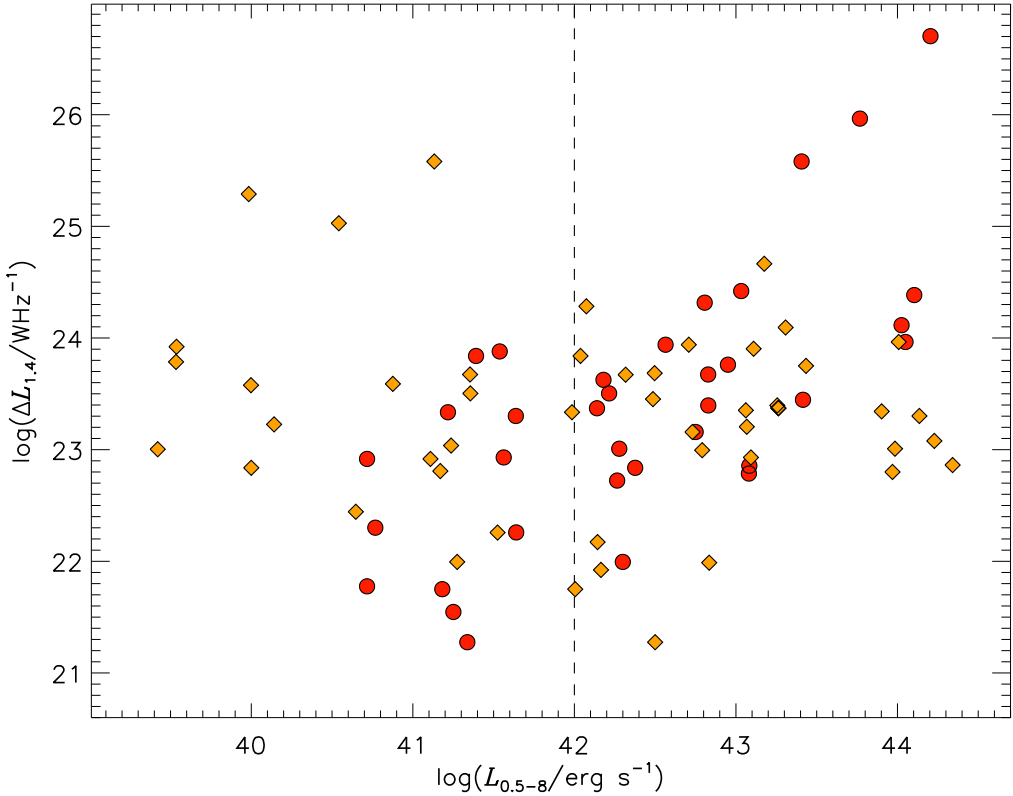, width=0.82\linewidth}
\caption{Excess radio luminosity above that predicted by the FIRRC as a function of X-ray luminosity for AGN-powered (red circles) and hybrid (orange diamonds) sources.}
\label{Fig:Radio_vs_X-ray}
\end{figure*}

The IRAC colour distribution of the sub-sample is shown in Figure \ref{Fig:irac_colours}. The AGN-powered and hybrid sources have a wide distribution in IRAC colour space and can lie well outside the region governed by the \citet{Donley:12} criteria. In Figure \ref{Fig:lrad_v_lir}, we plot rest-frame 1.4\,GHz luminosity versus star-forming total IR luminosity to show which sources follow the FIRRC. The AGN-powered sources, by definition, lie above the FIRRC because of the additional AGN component to radio emission. Hybrids on the other hand reside close to the FIRRC in a similar manner to the SF-powered sources. In Figure \ref{Fig:AGN_identified} we show the $q_{\rm IR}$ distribution as a function of 1.4\,GHz luminosity. The additional component to the radio luminosity for the AGN-powered sources results in lower $q_{\rm IR}$ values. In a similar manner to Table \ref{Table:AGN_identified}, we also signify how the AGN indicator results vary for each source. For instance, it is visible that many of the AGN identified through their X-ray emission lie on the FIRRC. For these particular sources we next plot their excess radio luminosity above that predicted by the FIRRC as a function of X-ray luminosity in Figure \ref{Fig:Radio_vs_X-ray}. We determine the Spearman rank  coefficient, $\rho$, and the significance of the deviation of $\rho$ from zero, $\sigma$, for the AGN-powered and hybrid sources. We find a moderate correlation for the AGN-powered sources ($\rho$\,=\,0.66, $\sigma$\,=\,10$^{-5}$) and no correlation for the hybrids ($\rho$\,=\,0.07, $\sigma$\,=\,0.55).

\begin{figure*}
\epsfig{file=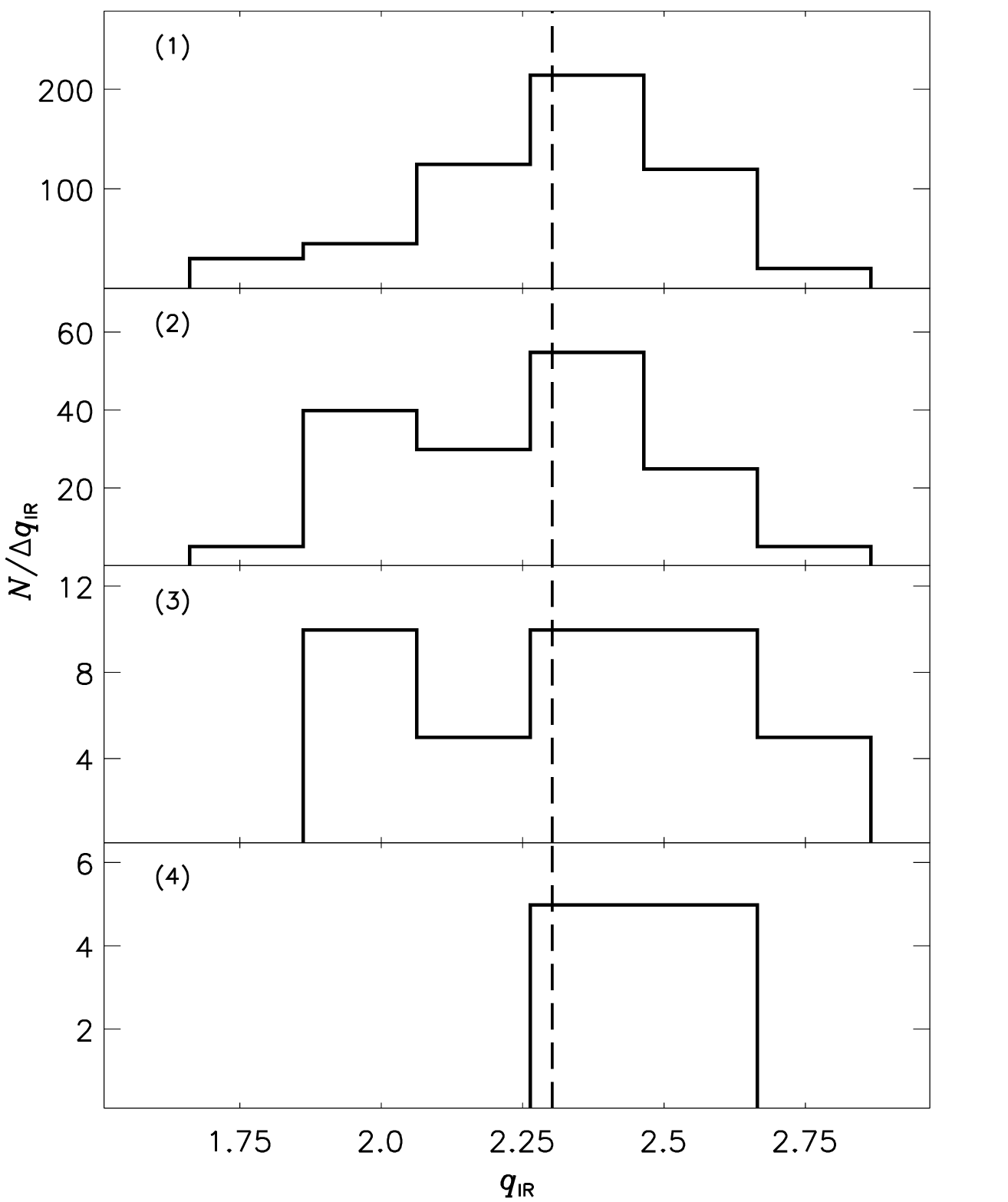, width=0.83\linewidth}
\caption{The $q_{\rm IR}$ distributions of SF-powered sources for different 1.4\,GHz flux density bins. The flux density bin ranges are as follows: (1) 0.030$-$0.067\,mJy; (2) 0.067$-$0.150\,mJy; (3) 0.150$-$0.333\,mJy; and (4) 0.333$-$0.742\,mJy. The dashed vertical line in each panel represents the mean $q_{\rm IR}$ value.}
\label{Fig:qir_dist_sf}
\end{figure*}

\begin{figure*}
\epsfig{file=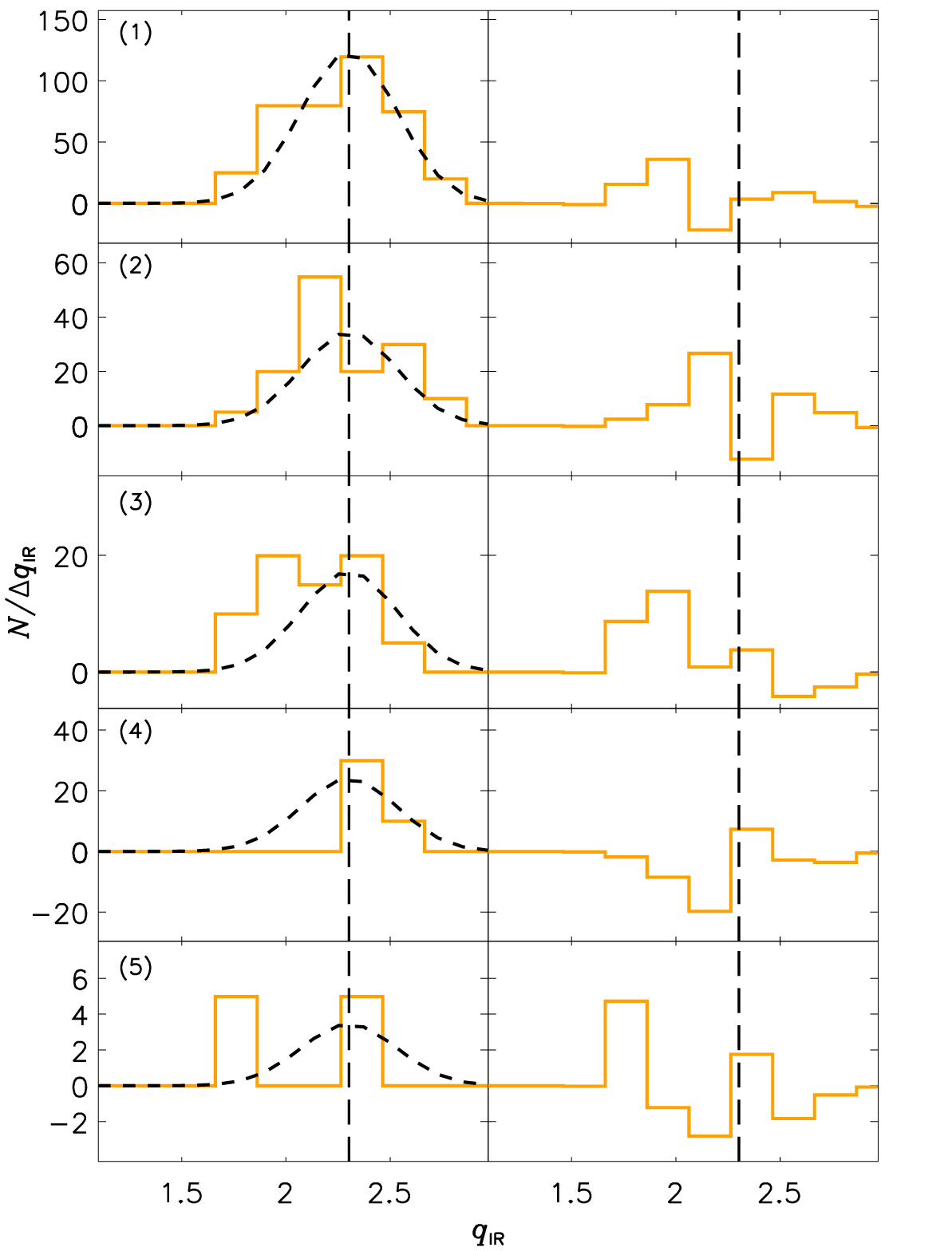, width=0.92\linewidth}
\caption{Left panels: The $q_{\rm IR}$ distributions of hybrid sources as solid orange lines and the normalised SF-powered Gaussian distributions as black dashed lines. Right panels: The residual distributions as solid orange lines, after the normalised SF-powered Gaussian distributions have been subtracted from the binned hybrid $q_{\rm IR}$ distributions. The various panels vertically show the $q_{\rm IR}$ and residual distributions for different 1.4\,GHz flux density bins. The flux density bin ranges are as follows: (1) 0.030$-$0.067\,mJy; (2) 0.067$-$0.150\,mJy; (3) 0.150$-$0.333\,mJy; (4) 0.333$-$0.742\,mJy; and (5) 0.742$-$1.653\,mJy. The dashed vertical line in each panel represents the mean $q_{\rm IR}$ value of the SF-powered sources.}
\label{Fig:qir_dist_hybrid}
\end{figure*}

Although it can be seen in Figures \ref{Fig:lrad_v_lir} and \ref{Fig:AGN_identified} that the hybrid sources have similar $q_{\rm IR}$ values to the SF-powered sources, we next investigate their $q_{\rm IR}$ distribution more quantitatively. Firstly, for four 1.4\,GHz flux density bins of equal size in log space, the number of SF-powered sources per $q_{\rm IR}$ bin width of 0.2 as a function of $q_{\rm IR}$ is shown in Figure \ref{Fig:qir_dist_sf}. The $q_{\rm IR}$ distribution of the SF-powered sources is close to a Gaussian distribution and these sources have a mean $q_{\rm IR}$ of 2.30 and a standard deviation of 0.23. This value for $q_{\rm IR}$ is consistent with those found by other studies ($q_{\rm IR}$\,=\,2.40\,$\pm$\,0.19, \citealt{Seymour:11}; $q_{\rm IR}$\,=\,2.40\,$\pm$\,0.24, \citealt{Ivison:10}; $q_{\rm IR}$\,=\,2.54\,$\pm$\,0.21, \citealt{Sajina:08}). The hybrid sources have a comparable mean $q_{\rm IR}$ of 2.26 and a standard deviation of 0.25. For five 1.4\,GHz flux density bins of equal size in log space, the number of hybrids per $q_{\rm IR}$ bin width of 0.2 as a function of $q_{\rm IR}$ is given in Figure \ref{Fig:qir_dist_hybrid}. If the same process is powering the radio emission in the hybrids, i.e. star formation, then it can be expected that their $q_{\rm IR}$ distribution would be Gaussian also and it would be spread uniformly around the mean $q_{\rm IR}$ of the SF-powered sources. On the other hand, if the radio emission is powered by an AGN for a significant fraction of these sources, then there should be a significant excess in the number of sources with a $q_{\rm IR}$ below the mean. For each flux density bin, to measure any potential excess as a function of $q_{\rm IR}$, firstly a Gaussian function with a mean and $\sigma$ taken from the SF-powered sources in the same flux density bin is normalised to the observed distribution of the hybrids. The normalisation is such that the integral of the Gaussian curve from the mean SF-powered $q_{\rm IR}$ out to infinity, is equal to the number of hybrids with a $q_{\rm IR}$ value above the mean SF-powered $q_{\rm IR}$. The normalised Gaussian curve is then integrated in each $q_{\rm IR}$ bin range and subtracted from the binned $q_{\rm IR}$ distribution to create a residual distribution. The right-hand panels of Figure \ref{Fig:qir_dist_hybrid} show that this residual distribution is small across all flux density bins. By summing the residual distribution across all $q_{\rm IR}$ and flux density bins, we find that there is an excess of 14 more hybrids with a $q_{\rm IR}$ below the mean SF-powered $q_{\rm IR}$ than above. However, this number is not significant as it is comparable to the Poisson uncertainty in the total number of hybrids in the sub-sample ($\sqrt{132}$\,$\approx$\,11).

We test the sensitivity of our results with respect to the FIRRC indicator by increasing the threshold such that a source is identified as an AGN only if it resides more than 5\,$\sigma$ away from the FIRRC. The overall number of sources found to contain an AGN decreases from 227 to 200 sources. The number of SF-powered and AGN-powered sources as a result of this change is 180 and 53, respectively. There are also now 147 hybrids. Across all radio flux density bins, there are now 21 more hybrids with a $q_{\rm IR}$ above the mean SF-powered $q_{\rm IR}$ than below. With a Poisson uncertainty in the number of hybrids of $\sqrt{147}$\,$\approx$\,12, this number is still not significant. Finally, the fractions of sources powered by star formation that host an AGN across the three redshift bins ($z$\,$<$\,0.5, 0.5\,$<$\,$z$\,$<$\,1 and 1\,$<$\,$z$\,$<$\,1.4) decrease by $<$\,2 per cent.

\begin{figure*}
\epsfig{file=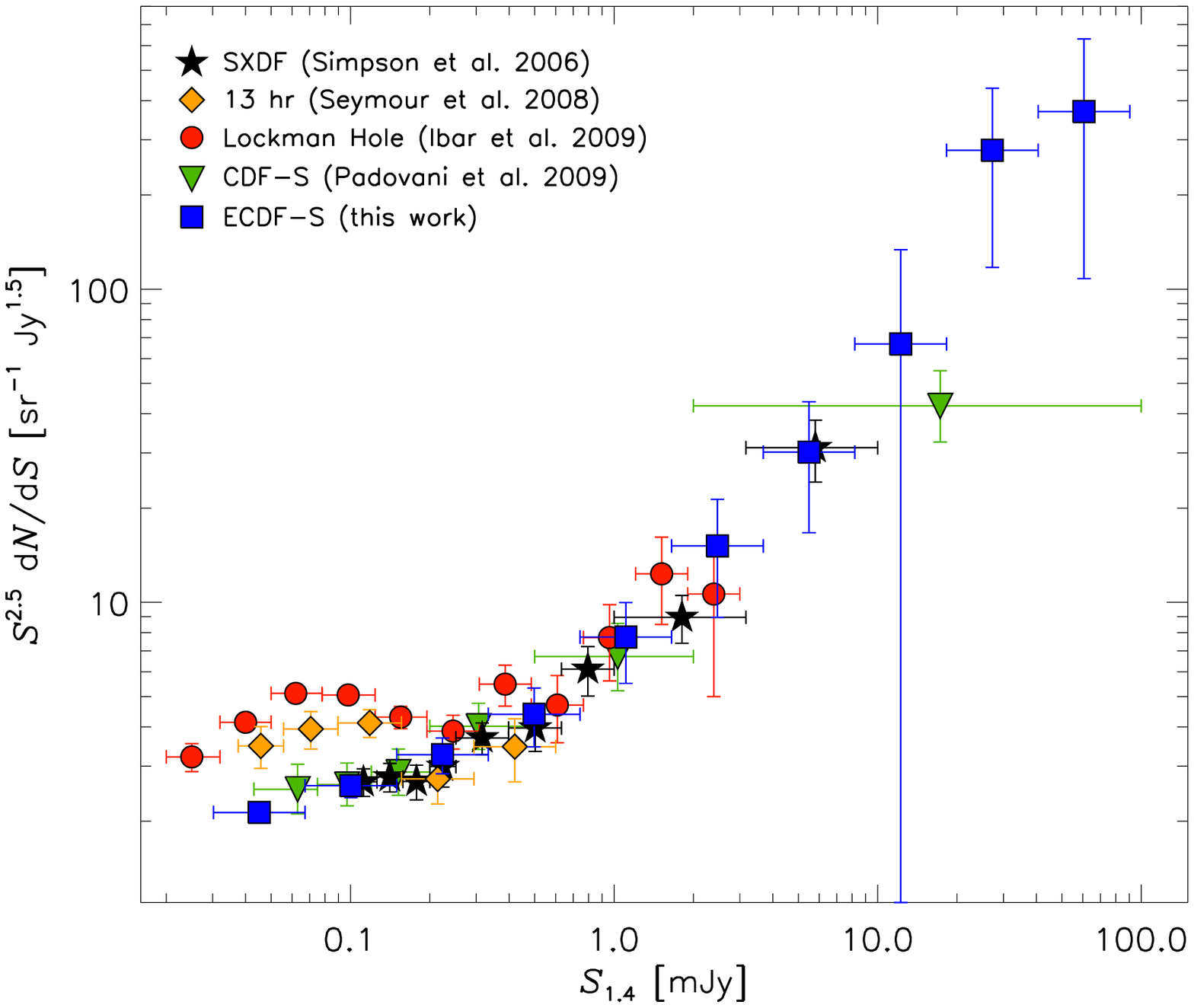, width=0.99\linewidth}
\caption{The Euclidean-normalised 1.4\,GHz source counts for the radio sources in the ECDF-S (blue squares). Also given are the source counts for the following fields: SXFD (\citealt{Simpson:06}, black stars); 13\,hr (\citealt{Seymour:08}, orange diamonds); Lockman Hole (\citealt{Ibar:09}, red circles); and CDF-S (\citealt{Padovani:09}, green upside down triangles).}
\label{Fig:Source_counts_all}
\end{figure*}

\begin{figure*}
\epsfig{file=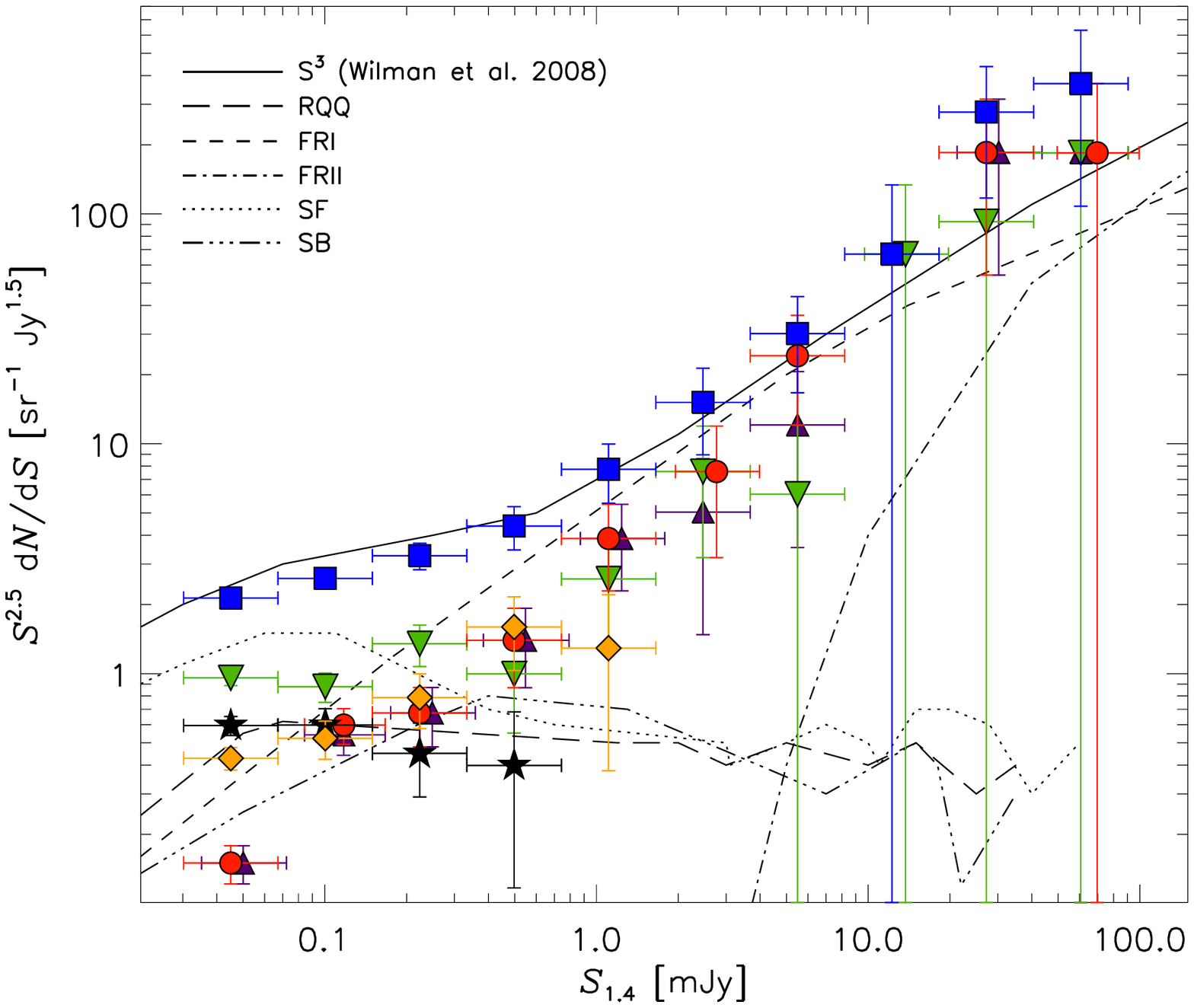, width=0.99\linewidth}
\caption{The Euclidean-normalised 1.4\,GHz source counts for the radio sources in the ECDF-S. Colours and symbols are as follows: total, blue squares; SF-powered, black stars; AGN-powered, red circles; hybrid, orange diamonds; `no-$z$', green upside down triangles; and LERGs, purple triangles. For clarity, some counts have been offset slightly along the abscissa. Also given are the source count models from the SKA simulations \citep{Wilman:08}.}
\label{Fig:Source_counts}
\end{figure*}

Next the Euclidean-normalised radio source counts are calculated for the ECDF-S, using the same radio flux density bins as for the $q_{\rm IR}$ distributions, with the extension to higher flux densities to incorporate the AGN-powered sources. For this calculation, we include the 266 sources in the full-sample that have unknown redshifts or have redshift uncertainties that were too large to be able to constrain the AGN and galaxy ISM contributions (see Section \ref{Section:Sample selection and redshifts}, hereafter `no-$z$ sources'). We show the Euclidean-normalised source counts for the ECDF-S in Figure \ref{Fig:Source_counts_all} along with the counts for other fields as a comparison. In Figure \ref{Fig:Source_counts} and Table \ref{Table:Source_counts}, we breakdown the source counts into the SF-powered, AGN-powered, hybrid and `no-$z$' components. Additionally in Figure \ref{Fig:Source_counts}, we show the LERG component of the AGN-powered source counts along with the models from the simulations for the Square Kilometre Array \citep[SKA,][]{Wilman:08}. Fractional uncertainties in the counts have been calculated by assuming a Poisson distribution and are equal to 1\,/\,$\sqrt{N}$, where $N$ is the number of sources in each bin. We have corrected for the reduced coverage of the ECDF-S observation at the fainter flux levels using Figure 2 from \citet{Miller:13}, which shows the fractional area covered by the survey as a function of sensitivity. When creating the sensitivity map, \citet{Miller:13} removed sources with peak flux densities greater than 150\,$\mu$Jy. It is assummed that along with the removal of point sources from the map, extended sources were removed also. We also corrected for the reduced coverage because of the footprints of observations at other wavelengths and for removing radio sources with inaccurate/contaminated photometry. We find that there is a sharp decrease in the overall source counts with decreasing flux density and a flattening below 1\,mJy. The AGN-powered sources dominate the source counts at bright flux densities and their counts decrease sharply at fainter flux densities with the form of a power law. The majority of other sources emerge only below $\sim$\,1\,mJy.

\section{Discussion}
\label{Section:Discussion}
The first thing to note is that many radio sources in the ECDF-S contain an AGN. Although this is expected at high flux densities, bright radio sources are relatively rare and so this result may be surprising for a sample dominated by faint sources. The IRAC colour AGN indicator and the criterion for an AGN component being required in the IR SED were both conservative. This has resulted in more reliable AGN detections compared to a scenario where these detection thresholds were lower. Furthermore, the effect of the FIRRC AGN indicator on the overall detection fraction was examined by forcing this indicator to be more conservative also. The effect was small. To compare individual indicator results to those of previous studies, the fraction of AGN in the sub-sample found through the radio spectral index indicator is 22 per cent (13/60 sources with derived spectral indices). This is significantly lower than the fraction found by \citet{Seymour:08} of 69 per cent (31/45). However, this is because there was no requirement for the full 3\,$\sigma$ confidence range of the spectral indices to lie outside the range for the SFGs in the \citet{Seymour:08} sample. Furthermore, the fraction of sources found to be an AGN through their X-ray emission (52 per cent) is comparable to the fraction found by \citet[hereafter P09]{Padovani:09} of 58 per cent for radio-selected sources, using the same X-ray luminosity cut.
 
We next discuss the IRAC colour distribution of the sub-sample. For most of the sources whose radio emission is powered by an AGN, it can be seen that they have blue IRAC colours which are consistent with old stellar populations. These objects are likely to be low-luminosity low-excitation radio galaxies (LERGs), hosted by ellipticals whose non-stellar light is weak \citep{Lacy:04, Gurkan:14}. LERGs do not have a strong ionising continuum and therefore they lack the hot dust required to produce the near- to mid-IR power law SED described by the \citet{Donley:12} criteria. On the other hand, the AGN-powered sources that do have a power law SED are the more conventional powerful radio-loud AGN such as quasars and broad-line radio galaxies (BLRGs) and the radio-quiet AGN which have enough radio emission from the AGN to lie off the FIRRC. Sources whose radio emission is powered by star formation predominantly have slightly redder colours and their place in IRAC colour parameter space is in agreement with that found by \citet{Hatziminaoglou:09}. 

The hybrid sources, by definition, are galaxies which host an AGN that does not clearly power the radio emission and therefore they consist of radio-quiet AGN. They have the most diverse IRAC colour distribution. The hybrids that cover the colour space associated with mid-IR-selected AGN are typical QSOs and Seyfert galaxies which have a strong ionising continuum, while those that have IRAC colours which are similar to SF-powered sources are likely to be less powerful radio-quiet AGN. A final region in the IRAC colour space of hybrid sources is one of blue $S_{\nu}$\,(5.8\,$\mu$m)\,/\,$S_{\nu}$\,(3.6\,$\mu$m) and very red $S_{\nu}$\,(8\,$\mu$m)\,/\,$S_{\nu}$\,(4.5\,$\mu$m) colours. This region typically consists of low-redshift obscured AGN with strong PAH features from a dominant host galaxy \citep{Lacy:04, Feltre:13}. It is reasonable that we find that the IRAC colour parameter space covered by the AGN-powered and hybrid sources extends well beyond the region described by the \citet{Donley:12} criteria. This result is because we have identified many AGN which do not have a dominant AGN component in their IR SED, and therefore do not have a near- to mid-IR power law. Since the hybrid sources follow the FIRRC, this means their AGN IR emission is significantly diluted by emission from star-forming processes. However, the AGN can still have an important contribution to the near- to mid-IR emission \citep{Hatziminaoglou:09}, particularly for the sources that lie within the IRAC colour region described by the \citet{Donley:12} criteria.

\begin{landscape}
\begin{table}
\begin{center}{
\caption{Euclidean-normalised 1.4\,GHz source counts. Columns are as follows: (1) flux density bin range; (2) central flux density of bin in log space; (3) total number of sources in bin; (4), (5) (6) and (7) number of AGN-powered, hybrid, SF-powered and `no-$z$' sources in each bin, respectively; (8) multiplicative correction factor due to the reduced coverage as described in Section \ref{Section:Results}; (9) Corrected effective area, $A_{\rm eff}$; (10) total source counts; and (11), (12), (13) and (14) AGN-powered, hybrid, SF-powered and `no-$z$' source counts, respectively.}
\begin{tabular}{@{}cccccccccccccc@{}}
\hline
\hline
$S_{\rm \nu}$ range & Central $S_{\rm \nu}$ & \multicolumn{5}{c}{No.} & Correction & $A_{\rm eff}$ & \multicolumn{5}{c}{Counts}\\
& & Total & AGN & Hybrid & SF & No$-z$ & & & Total & AGN & Hybrid & SF & No$-z$\\
\,[mJy] & [mJy] & & & & & & & [deg$^{2}$] & [sr$^{-1}$\,Jy$^{1.5}$] & [sr$^{-1}$\,Jy$^{1.5}$] & [sr$^{-1}$\,Jy$^{1.5}$] & [sr$^{-1}$\,Jy$^{1.5}$] & [sr$^{-1}$\,Jy$^{1.5}$]\\
(1) & (2) & (3) & (4) & (5) & (6) & (7) & (8) & (9) & (10) & (11) & (12) & (13) & (14)\\
\hline
0.030$-$0.067 & 0.045 & 398 & 28 & 80 & 111 & 179 & 1.30 & 0.23 & 2.13\,$\pm$\,0.11         & 0.15\,$\pm$\,0.03 & 0.43\,$\pm$\,0.05 & 0.59\,$\pm$\,0.06 & 0.96\,$\pm$\,0.07\\
0.067$-$0.150 & 0.100 & 139 & 32 & 28 & 32 & 47 & 1.50 & 0.21     & 2.60\,$\pm$\,0.22         & 0.60\,$\pm$\,0.11 & 0.52\,$\pm$\,0.10 & 0.60\,$\pm$\,0.11 & 0.88\,$\pm$\,0.13\\
0.150$-$0.333 & 0.223 & 58 & 12 & 14 & 8 & 24 & 1.36 & 0.24         & 3.26\,$\pm$\,0.43         & 0.67\,$\pm$\,0.19 & 0.79\,$\pm$\,0.21 & 0.45\,$\pm$\,0.16 & 1.35\,$\pm$\,0.28\\
0.333$-$0.742 & 0.497 & 22 & 7 & 8 & 2 & 5 & 1.45 & 0.22               & 4.39\,$\pm$\,0.94         & 1.40\,$\pm$\,0.53 & 1.60\,$\pm$\,0.56 & 0.40\,$\pm$\,0.28 & 1.00\,$\pm$\,0.45\\
0.742$-$1.653 & 1.108 & 12 & 6 & 2 & 0 & 4 & 1.42 & 0.23               & 7.75\,$\pm$\,2.24         & 3.87\,$\pm$\,1.58 & 1.29\,$\pm$\,0.91 & 0.00                        & 2.58\,$\pm$\,1.29\\
1.653$-$3.680 & 2.466 & 6 & 3 & 0 & 0 & 3 & 1.67 & 0.19                 & 15.14\,$\pm$\,6.18       & 7.57\,$\pm$\,4.37 & 0.00 & 0.00                                                & 7.57\,$\pm$\,4.37\\
3.680$-$8.194 & 5.491 & 5 & 4 & 0 & 0 & 1 & 1.20 & 0.27                 & 30.19\,$\pm$\,13.50     & 24.15\,$\pm$\,12.07 & 0.00 & 0.00                                            & 6.04\,$\pm$\,6.04\\
8.194$-$18.244 & 12.227 & 1 & 0 & 0 & 0 & 1 & 4.00 & 0.08             & 66.86\,$\pm$\,66.86     & 0.00 & 0.00 & 0.00                                                                       & 66.86\,$\pm$\,66.86\\
18.244$-$40.623 & 27.224 & 3 & 2 & 0 & 0 & 1 & 1.67 & 0.19           & 277.68\,$\pm$\,160.32 & 185.12\,$\pm$\,130.91 & 0.00 & 0.00                                        & 92.56\,$\pm$\,92.56\\
40.623$-$90.450 & 60.616 & 2 & 1 & 0 & 0 & 1 & 1.00 & 0.32           & 369.04\,$\pm$\,260.95 & 184.52\,$\pm$\,184.52 & 0.00 & 0.00                                        & 184.52\,$\pm$\,184.52\\
\hline
\end{tabular}

\label{Table:Source_counts}}
\end{center}
\end{table}
\end{landscape}

Regarding the star-forming total IR luminosities of the sub-sample, the greater fractions of NIRGs and LIRGs over ULIRGs is simply down to more luminous sources being rarer. We compare the fractions of NIRGs, LIRGs and ULIRGs to those for the \citet{Symeonidis:13} sample of SF-powered, IR-luminous galaxies at similar redshifts ($z$\,$\lesssim$\,2). The fraction of NIRGs in our sub-sample of 43 per cent is higher than that for the \citet{Symeonidis:13} sample (20 per cent). We also find lower fractions of LIRGs and ULIRGs (53 per cent and 4 per cent, respectively) than the fractions of 64 per cent and 20 per cent for the \citet{Symeonidis:13} sample. This is reasonable because our sub-sample is not IR-selected and some sources have been selected because of their radio emission from an AGN. Furthermore, SF-powered sources must be highly star-forming to be detected in the radio survey, but this is not the case for AGN-powered sources. It is because of this selection effect that higher SFRs are observed in the SF-powered sources compared to the AGN-powered sources. The low SFRs that we derive for the AGN-powered sources are in agreement with those that have been observed in other low-redshift AGN hosts \citep[$\langle$SFR$\rangle$\,=\,18$-$41\,M$_{\rm \odot}$\,yr$^{-1}$, 0.4\,$<$\,$z$\,$<$\,0.9,][]{Seymour:11}. 

The fraction of sources powered by star formation that host an AGN has been observed to decrease with redshift. At $z$\,$<$\,0.5, the majority of star-forming sources in the sub-sample host an AGN while at $z$\,$>$\,0.5, only a minority of sources do. However, this observation is a result of the fact that for high-redshift sources, the IRAC photometry were not used during the SED fitting because they measure the emission at $<$\,2\,$\mu$m (rest-frame). Without the IRAC photometry, the AGN component cannot be well constrained and thus only low-redshift AGN are identified by the SED fitting technique. If the results of the SED fitting are not considered then the fraction of star-forming sources at $z$\,$<$\,0.5 that host an AGN drops to 0.05. 

Regarding the $q_{\rm IR}$ distribution of the hybrids, there was no significant excess in the number of sources with a $q_{\rm IR}$ below the mean $q_{\rm IR}$ of the SF-powered sources. While on an individual basis the radio emission of a hybrid source may be AGN-powered, the overall $q_{\rm IR}$ distribution indicates that the majority of hybrids are SF-powered in the radio. Since the hybrid sources are in essence radio-quiet AGN, this result supports previous evidence for the radio emission from radio-quiet AGN being predominantly a result of star-forming processes \citep[e.g.][]{Kimball:11, Padovani:11, Padovani:14}. Focusing on the sub-group of hybrids with an AGN identified through their X-ray emission, finding no correlation between excess radio and X-ray luminosities is further evidence that the AGN in the hybrids is not the source of their radio emission. The excess radio luminosity distribution is likely to be a result of residual scatter after the subtraction of the emission that is predicted by the FIRRC. Studies into the radio and IR emission of radio-quiet Seyfert galaxies \citep[e.g.][]{Roy:98} have found them to follow the FIRRC. Seyfert galaxies constitute some of the X-ray luminous hybrids and therefore our result is consistent with these previous findings also.
 
The flux density at which a flattening of the Euclidean-normalised source counts is observed is comparable to that for the source counts of other surveys, e.g. the 13\,hr\,{\it XMM-Newton/Chandra Deep Field} \citep{Seymour:08}, Subaru/{\it XMM-Newton} Deep Field \citep[SXDF,][]{Simpson:06}, Lockman Hole \citep{Ibar:09} and a smaller area of the CDF-S covered by P09. The total source counts of the ECDF-S are in agreement with the counts from \citet{Simpson:06} and P09. Below $\sim$0.2\,mJy they are slightly lower than those from \citet{Seymour:08} and \citet{Ibar:09}. It has been shown that while the counts from various radio surveys are consistent at most flux densities, below 1\,mJy there are some discrepancies \citep{Simpson:06, Condon:12, Heywood:13}.

Along with calculating the radio source counts for other fields, other studies have also separated the counts by galaxy type. For example, \citet{Seymour:08} separated the radio sources in the 13\,hr field into SFGs and AGN, although they only used radio indicators and so they did not differentiate between SFGs and sources that contained an AGN that did not power the radio emission. P09 on the other hand, did use non-radio indicators and calculated a radio-quiet component of their AGN source counts. Excluding this radio-quiet component, P09 found an AGN contribution to the source counts of $\sim$10 per cent in their faintest bin (43$-$75\,$\mu$Jy). \citet{Seymour:08} found a slightly higher AGN contribution of $\sim$30 per cent in their faintest bin (38$-$56\,$\mu$Jy). We find a contribution of 10$-$50 per cent from the AGN-powered sources in the fainest bin of this study (30$-$67\,$\mu$Jy), consistent with both studies above. The uncertainty in this contribution incorporates the uncertain nature of the `no-$z$' sources in the full-sample. We find a contribution of 30$-$70 per cent from the SF-powered sources to the counts in our faintest bin, in agreement with the contribution from P09 of $\sim$60 per cent. Regarding the hybrids in this study, we find they contribute 20$-$65 per cent in our faintest bin, which is consistent with the radio-quiet AGN contribution of 15$-$40 per cent from P09. The contribution we find for the hybrid sources is also in agreement with the fraction of radio-quiet AGN found by \citet{Bonzini:13} for the same radio observations. Increasing the FIRRC AGN indicator threshold to 5\,$\sigma$ effects the hybrid contribution by $<$\,5 per cent. Therefore, increasing this threshold has only a small effect on this result. At higher flux densities, the hybrid contribution remains approximately constant up to $\sim$1\,mJy and the hybrid and P09 radio-quiet AGN source counts are consistent at 3\,$\sigma$. Using X-ray data, \citet{Simpson:06} also found that radio-quiet AGN are a significant fraction of faint ($S_{\rm 1.4}$\,$<$\,300\,$\mu$Jy) radio sources. If we sum the SF-powered and hybrid components, the total contribution to the counts of 50$-$90 per cent in the faintest bin is in agreement with the SFG contribution from \citet{Seymour:08} of $\sim$70 per cent. Overall, our findings suggest that galaxies which host an AGN but whose radio emission is likely powered by star formation play a significant role at faint flux densities. Furthermore, they have a significant contribution to the flattening of the source counts that is observed for the extragalactic radio source population.

Comparing our source counts to the SKA simulations as shown in Figure \ref{Fig:Source_counts}, we find our total counts are consistent with those predicted at 1.4\,GHz. We find that LERGs, the AGN-powered objects which do not have a strong ionising continuum and therefore lie outside the \citet{Donley:12} region in IRAC colour space, dominate the AGN-powered source counts. Using a different classification system, \citet{Wilman:08} predict that the less-luminous Fanaroff-Riley type I (FR I) class dominates over the more powerful type II (FR II) class below 50\,mJy. A correlation between optical emission line and radio luminosities has been observed for nearby radio-loud AGN \citep{Kauffmann:08}, albeit with a large scatter, indicating that LERGs are often FR Is. Our results are therefore in general agreement with the radio-loud AGN models from the SKA simulations. Furthermore, above 0.1\,mJy the power law slope for the LERGs is consistent with the slope for the FR Is. One inconsistency between our results and the SKA simulations, however, is that the hybrid contribution to the counts is approximately half an order of magnitude greater than that predicted for radio-quiet AGN. With regards to the power law of the overall source counts of AGN-powered objects, through modelling the evolution of the radio luminosity function, \citet{Seymour:04} showed that this extends to flux densities below $\sim$\,50\,$\mu$Jy. The SFG component, however, is much flatter in the $\sim$\,50\,$\mu$Jy$-$1\,mJy range and the extrapolation of the SFG model predicts that at a flux density of $\sim$10\,$\mu$Jy, the SFG source counts will be greater than the AGN counts by an order of magnitude. To fainter flux densities, these counts will then decrease in a similar manner to the way AGN counts decrease from $\sim$100\,mJy due to the expansion of the Universe. \citet{Wilman:08} and \citet{Bethermin:12} both showed that a significant fraction of the SFG population at $\sim$\,50\,$\mu$Jy are secularly-evolving galaxies. Note that a radio-quiet component was not included in the latter source count models. Future studies which probe the sub-50\,$\mu$Jy radio source population are required in order to test these models at these flux densities and to investigate the nature of even fainter AGN. \citet{Jarvis:04} modelled the evolution of the radio-quiet component of the radio luminosity function and found that radio-quiet AGN begin to significantly contribute to the source counts at $\sim$\,0.5\,mJy. A prediction which has been confirmed by later observational studies \citep[e.g.][]{Simpson:06}. Future studies modelling the radio counts for the different source populations below $\sim$50\,$\mu$Jy will need to present a scenario which depicts a strong contribution from radio-quiet AGN.

\section{Conclusion}
\label{Section:Conclusion}
We have fitted the IR SEDs of 380 radio sources in the ECDF-S at $z$\,$\lesssim$\,1.4 with models that have ISM and AGN components. This has enabled us to separate the IR emission from star formation and an AGN in these sources. With the additional use of various multi-wavelength AGN indicators, our results have revealed that the majority of sources contain an AGN. Many of the AGN indicators used in this study are conservative and therefore we can be confident that the AGN identifications are reliable. We determined that for 25 per cent of these sources, the radio emission is powered by an AGN. For 40 per cent, there is no sign of an AGN and their radio emission is therefore being powered by star formation. The remaining 35 per cent are hybrids in the sense that they contain an AGN but their star formation is likely to be powering the radio emission. The implication of these results is that radio sources which have likely been selected on their star formation have a high AGN fraction. 

From the ISM component of the fitted SEDs, some high-redshift sources have total IR luminosities that correspond to extreme SFRs ($\gtrsim$\,1000\,M$_{\rm \odot}$\,yr$^{-1}$). Such high SFRs are also observed in {\it Herschel}-selected and non-{\it Herschel}-selected galaxies at higher redshifts than in this study. For the most part however, the SFRs of the sub-sample are much lower than 1000\,M$_{\rm \odot}$\,yr$^{-1}$, with a median SFR of 20\,M$_{\rm \odot}$\,yr$^{-1}$ and a median absolute deviation of 20\,M$_{\rm \odot}$\,yr$^{-1}$.

AGN-powered sources dominate the Euclidean-normalised source counts at radio flux densities above 1\,mJy. At fainter flux densities, a flattening of the total source counts is observed and the contributions from SF-powered and hybrid sources become significant. At the faintest 1.4\,GHz flux density bin in this study of $\sim$50\,$\mu$Jy, we find that the hybrid sources contribute 20$-$65 per cent to the total source counts. Therefore, they have a significant contribution, along with star-forming sources which do not host an AGN, to the flattening of the counts that is observed below 1\,mJy for the extragalactic radio source population.

\section{Acknowledgments}
We thank the anonymous referee for their comments which improved the manuscript. We also thank Neal Miller for useful discussions and James Mullaney for help regarding {\tt DecompIR}. JIR acknowledges the support of a Science and Technologies Facilities Council studentship. MJP acknowledges support from the Science and Technology Facilities Council (STFC) [grant number ST/K000977/1]. SJO and MS acknowledge support from the STFC [grant number ST/I000976/1]. NS is the recipient of an Australian Research Council Future Fellowship. MV acknowledges support from the South African Department of Science and Technology (DST/CON 0134/2014), the European Commission Research Executive Agency (FP7-SPACE-2013-1 GA 607254) and the Italian Ministry for Foreign Affairs and International Cooperation (PGR GA ZA14GR02). The Dark Cosmology Centre is funded by the Danish National Research Foundation. SPIRE has been developed by a consortium of institutes led by Cardiff Univ. (UK) and including: Univ. Lethbridge (Canada); NAOC (China); CEA, LAM (France); IFSI, Univ. Padua (Italy); IAC (Spain); Stockholm Observatory (Sweden); Imperial College London, RAL, UCL-MSSL, UKATC, Univ. Sussex (UK); and Caltech, JPL, NHSC, Univ. Colorado (USA). This development has been supported by national funding agencies: CSA (Canada); NAOC (China); CEA, CNES, CNRS (France); ASI (Italy); MCINN (Spain); SNSB (Sweden); STFC (UK); and NASA (USA). This work is also based on observations made with the {\it Spitzer Space Telescope}, which is operated by the Jet Propulsion Laboratory (JPL), California Institute of Technology (Caltech) under contract with NASA. This work benefitted from the NASA/IPAC Extragalactic Database (NED), which is operated by the JPL, Caltech, under contract with NASA.

\bibliographystyle{mnras}
\bibliography{AGN_06_references}
\bsp

\label{lastpage}

\end{document}